\documentclass[graybox]{svmult}

\usepackage{mathptmx}       % selects Times Roman as basic font
\usepackage{helvet}         % selects Helvetica as sans-serif font
\usepackage{courier}        % selects Courier as typewriter font
\usepackage{type1cm}        % activate if the above 3 fonts are
\usepackage{makeidx}         % allows index generation
\usepackage{graphicx}        % standard LaTeX graphics tool
\usepackage{multicol}        % used for the two-column index
\usepackage[bottom]{footmisc}% places footnotes at page bottom
\usepackage{hyperref}        %for hyperlinks
\usepackage{soul}            % for high-lighting of text
\hypersetup{colorlinks=true}
\usepackage[square,numbers]{natbib}
\usepackage{aas_macros}
\usepackage{amsmath}
\usepackage{amssymb}
\newcommand{\dd}{{\rm d}}
\newcommand{\omegam}{\varOmega_{\rm m}}
\newcommand{\omegab}{\varOmega_{\rm b}}
\newcommand{\omegal}{\varOmega_{\varLambda}}
\newcommand{\omegak}{\varOmega_{\rm k}}
\newcommand{\omegar}{\varOmega_{\rm r}}
\newcommand\ion[2]{\text{#1\,\textsc{\lowercase{#2}}}}	% ionization states

\makeindex             % used for the subject index
\begin{document}
%\tableofcontents{}
\title*{X-ray cluster cosmology}
\author{Nicolas Clerc\thanks{corresponding author} and Alexis Finoguenov\thanks{corresponding author}}
% \thanks{corresponding author}
\institute{Nicolas Clerc \at IRAP, Universit{\'e} de Toulouse, CNRS, UPS, CNES, Toulouse, France\\ \email{nicolas.clerc@irap.omp.eu}
\and Alexis Finoguenov \at Department of Physics,
              Gustaf H{\"a}llstr{\"o}min katu 2 A,
              University of Helsnki, Helsinki, Finland\\ \email{alexis.finoguenov@helsinki.fi}}
\maketitle
\abstract{Formation of dark matter halos is sensitive to the expansion rate of the Universe and to the growth of structures under gravitational collapse. Virialization of halos heats the gaseous intra-cluster medium to high temperatures, leading to copious emission of photons at X-ray wavelengths. We summarize the progress of X-ray surveys in determining cosmology using galaxy clusters. We review recent cosmological results based on cluster volume abundance, clustering, standard candles, extreme object statistics, and present relevant theoretical considerations. We discuss clusters as gravitation theory probes and present an outlook on future developments.}

\section*{Keywords} 
X-rays, Clusters of galaxies, Cosmology, Large-scale structure of the universe, X-ray surveys, Dark matter, Dark energy, Halo mass function

\section{Introduction: Role of X-rays in cluster cosmology}

X-ray emission from clusters is predominantly of thermal origin, a convincing observational proof being the presence of atomic emission lines on top of bremsstrahlung continuum in X-ray spectra. The X-ray emitting gas, generically identified to the intra-cluster medium (ICM), fills the entire volume of massive, Mpc-sized clusters, within which hundreds to thousands of galaxies orbit under the influence of the gravitational force. When gas and galaxies are in the same gravitational potential, and galaxies are in virial equilibrium within the cluster potential, assuming gas particles have roughly similar three-dimensional velocity dispersion as galaxies\footnote{Numerical simulations show that galaxies move slightly supersonically through the ICM \citep{2005Faltenbacher}.}:
\begin{equation}
    k_B T/{\mu m_p} \sim \sigma_r^2 \sim G M_{\rm tot}/3R
\end{equation}
where $M_{\rm tot}$ and $R$ are typical mass and radius for a cluster, $k_B$ the Boltzmann constant, $m_p$ a proton mass and $G$ the Universal constant of gravitation.
Considering galaxy radial velocity dispersion $\sigma_r \simeq 1000$\,km\,s$^{-1}$ and mean molecular weight $\mu \simeq 0.6$ provides a temperature $T \simeq 7\times 10^7$\,K, equivalently $k_B T \simeq 6$\,keV. The gas with electron density $n_e \simeq 10^{-3}$\,cm$^{-3}$ is heavily ionized at this temperature. Thermal bremsstrahlung is the main X-ray emission mechanism in hot clusters, whose emissivity scales with the product of $n_p n_e$, with the proton density $n_p \simeq n_e/1.2$. Also, the plasma is optically thin to the emitted photons\footnote{Notable exception from this picture is driven by resonant absorption lines in cluster cores \citep[][]{1987Gilfanov}.}, so the black body spectrum does not form. The number of emitted photons is much lower than the number of baryons, and the emission is called diffuse, similar to supernova remnants, but different from compact sources where photon ionization is important. Outside of the cluster virial radius, where the density drops, in the presence of background UV radiation field, the ionization processes start to play a role \citep{2019KhabibullinChurazov}.
At temperatures below 1 keV processes driving line emission become relatively more important and most of them also scale with $n_p n_e$ (e.g.~bound-bound transitions due to collisional excitation).

The total X-ray luminosity of the gas scales with the square of density times the volume $V$; while the total gas mass scales with $n_p V$. Based on these arguments we expect total mass to correlate strongly with gas temperature, gas mass and X-ray luminosity, all of them are retrievable from the X-ray spectra, along with metallicity, which controls the contribution of the line emission to the total X-ray luminosity, described above. The importance of X-rays for cosmology can be understood from this scaling. Massive virialized halos are easily identified as luminous X-ray extended sources and their masses are readily available from X-ray measurements, once the source distance is known. The distance is obtained from a measurement of the cluster redshift. Redshift is most commonly inferred from optical observations of cluster galaxy members\footnote{Resolved emission lines in X-ray spectra may also provide a redshift measurement \citep[][]{2011YuTozzi, 2014Borm}.}. Furthermore, line-of-sight projection effects are low thanks to the emissivity scaling with $n^2$.

Detection of the cluster emission in X-ray requires space instrumentation due to opacity of the Earth atmosphere, and strongly depends on the energy band: extremely soft X-rays with energies $<0.3$ keV are strongly absorbed by the Milky Way halo and interstellar medium, and local foregrounds are high, so it is a relatively poorly understood regime, but important for low mass halos. The standard band for cluster detection is 0.5--2.0 keV, employed by \emph{ROSAT}, \emph{Chandra}, \emph{XMM-Newton}, \emph{SRG}/eROSITA studies. Slight variations of the choice of the band in order to increase the sensitivity of the survey are often made, as current X-ray instrumentation records\footnote{Albeit with low resolution - improvements involve the use of calorimeters, e.g. X-IFU \citep[][]{2020Barret}.} the energy of the detected photon and bands are not pre-set, unlike in the optical/IR photometry. At energies above 2-2.5 keV, the reflectivity of the mirrors drops (and off-axis reflection efficiency drops even more) with the exact behaviour dependent on the material (with leading performance of the gold coating), which reduces the collecting area of the X-ray telescope by an order of magnitude \citep[][and Section ``Optics for X-ray Astrophysics'' in this handbook]{1990Zombeck}. In addition, the emission of high-redshift clusters as well as galaxy groups is soft, due respectively to redshifting of the spectrum and to lower gas temperature resulting in thermal cut-off of the emission.

In order to be a competitive cosmological probe, detection of clusters has to be reproducible, robust, and sensitive to cosmology. Sensitivity to cosmology is due to scaling relations between X-ray properties and cluster total mass (for more details on this read Chapter ``Scaling relations of clusters and groups, and their evolution''). The actual observing tests use X-ray luminosity function of clusters, temperature function, as well as combination of gas mass and temperature (aka $Y_X$). The next sections provide multiple references to these works.

Collecting area of X-ray telescopes has been the largest challenge in the feasibility of cluster detection, due to inefficiency of X-ray reflection, requiring incidence angles close to zero, which correspondingly zeros out the mirror collecting area. The collecting area of current instrumentation is $\sim 1000$ cm$^2$, while in the future a gain by an order of magnitude is expected from \emph{Athena}. The sky background in the 0.5--2 keV is $10^{-15}$\,ergs\,s$^{-1}$\,cm$^{-2}$\,arcmin$^{-2}$, and the all-sky sensitivity achieved by eROSITA is $10^{-14}$\,ergs\,s$^{-1}$\,cm$^{-2}$\,arcmin$^{-2}$. Deeper fields observations, performed by \emph{Chandra} and \emph{XMM-Newton} reach $10^{-16}$\,ergs\,s$^{-1}$\,cm$^{-2}$\,arcmin$^{-2}$. The cluster size to cluster flux relation, as well as a confusion limit, implies that instrument resolution has to be better than arcminute at depths (in units ergs\,s$^{-1}$\,cm$^{-2}$) below $10^{-12}$, $<30^{\prime\prime}$ below   $10^{-13}$, $<10^{\prime\prime}$ below   $10^{-14}$, $<5^{\prime\prime}$ below $10^{-15}$, reaching 1 $^{\prime\prime}$ at $10^{-17}$ ergs s$^{-1}$ cm$^{-2}$ arcmin$^{-2}$. So, low resolution telescopes can only effectively perform shallow surveys, while for high-resolution telescopes, like \emph{Chandra}, only Msec exposures open up unique advantages. The number of clusters per square degree scales from 0.3 for \emph{ROSAT} All-Sky Survey (RASS) \citep{2020Finoguenov} to 3 for eROSITA \citep{2021LiuEFEDS}, 15 for wide \emph{XMM-Newton}/\emph{Chandra} surveys \citep{2017Pierre}, 100-1000 for deep and ultra-deep \emph{XMM} surveys \citep{2007Finoguenov, 2015Finoguenov}. 

For comparison, red sequence cluster search yields 5--10 clusters per square degree, depending on the depths of optical/NIR imaging \citep{2014Rykoff, 2016Rykoff}. Spectroscopic catalog of galaxy groups yield 10-1000 systems per square degree \citep{2005Gerke, 2012Knobel, 2017Tempel}, but large area surveys are still limited to $z<0.1$. 
The main bottleneck of optical surveys are the projection effects, which limits the mass range accessible to red sequence studies and overmerging/oversplitting problem affecting spectroscopic galaxy groups as well as developing an estimator of group mass, which in most algorithms comes with large scatter \citep{2015Old, 2018Old}. The strength of X-rays is to improve on the purity and completeness with respect to optical catalogs. As X-ray emission traces the virialization of gas, it characterizes the central part of the group, typically extended to the scale radius\footnote{See Sect.~\ref{sect:basic_cosmology} for the definition of $R_{500}$.} $R_{500}$, while efficient galaxy group search using spectroscopic surveys requires one to consider much larger radii. As a result, not all optical groups are supposed to be detected at X-rays, but only those with masses within $R_{500}$ above X-ray sensitivity limits. The open issue of effect of Active Galactic Nuclei (AGN) feedback on the intergalactic medium (IGM) further reduces our ability to predict X-ray properties of optical groups. Nevertheless, the cosmology using X-ray galaxy groups yields consistent results to clusters, which indicates that the prospects of understanding galaxy groups through X-rays are bright.

\section{Role of massive halos in cosmology\label{sect:basic_cosmology}}

Numerical simulations are generally required for a complete and detailed theoretical modelling of the formation of dark matter halos. However a `first order' treatment highlights many of the features relevant to cluster cosmology, such as the exponential decrease in the abundance of high-mass objects and the roles of matter-energy components of the cosmological model on the distribution of halos. This section provides a few elements of the theory of structure formation. By no means it is a comprehensive overview of the topic, which may be found in suitable cosmology textbooks. We take this opportunity to introduce some key parameters and notations frequently referred to in this Chapter.

\subsection{The homogeneous model}

The homogeneous, isotropic background as described by the Friedmann-Lema{\^i}tre-Robertson-Walker metric involves the dimensionless scale factor $a(t) = (1+z)^{-1}$, with $z$ representing the redshift and $t$ a universal cosmological time coordinate. Its evolution is governed by Einstein field equations linking curvature in the metric to the energy-momentum tensor, involving in particular the fundamental constants $G$ (Newton gravitation constant) and $c$ (speed of light in the vacuum). This tensor encapsulates contributions from non-relativistic matter with energy density $\rho_{\rm m}$ much greater than pressure $p_{\rm m}/c^2$; and from relativistic species (including radiation) of density $\rho_{\rm r} = 3 p_{\rm r}/c^2$. As a consequence of field equations, the Hubble parameter $H(a) = \dd (\ln a) / \dd t = \dot{a}/a$ for a model with cosmological constant\footnote{Equivalently, a component of density $\rho_{\varLambda} = \varLambda c^2/(8 \pi G)$.} $\varLambda$ writes:
\begin{equation}\label{eq:hubble_z}
    \frac{H(a)}{H_0} = E(z) = \left( \omegam a^{-3} + \omegar a^{-4} + \omegak a^{-2} + \omegal \right)^{1/2}
\end{equation}
with $\varOmega_i = \rho_i/\rho_{c}$ the present-day density of each component $i$, expressed relative to the critical density $\rho_{c}=3 H_0^2/(8 \pi G) \simeq 1.9 \times 10^{-29} h^2$\,g\,cm$^{-3}$, and $H_0 = H(z=0) = 100 h$\,km\,s$^{-1}$\,Mpc$^{-1}$ the present-day Hubble constant. This Chapter discusses structure formation at epochs well after the early Universe, one therefore neglects the relativistic component density. The curvature parameter $\omegak = 1-\omegam-\omegal$ vanishes in a flat Universe, although some of the cosmological tests using clusters are sensitive to deviations from flatness.

Equation~\ref{eq:hubble_z} applies to the Lambda Cold Dark Matter model ($\varLambda$CDM). The formulation is slightly modified when accounting for a Dark Energy equation of state $p_{\rm X}=w\rho_{\rm X} c^2$.
The case of a cosmological constant is recovered with $w=-1$. Conservation of the energy-momentum tensor implies $\rho_i(z) \propto (1+z)^{3(1+w_i)}$ for each of its components with constant equation of state parameter $w_i$. Furthermore the parameter $w$ may vary with redshift; a standard, although not unique, parametrization of the Dark Energy equation of state reads $w(a) = w_0 + w_1 (1-a)$.
The evolution factor in this case writes:
\begin{align}
E(z)^2 = \omegam (1+z)^3 &+ \omegar (1+z)^4 + \omegak (1+z)^2 \nonumber\\
&+ \varOmega_{\rm X} (1+z)^{3(1+w_0+w_1)}\; e^{-3 w_1 z/(1+z)}\label{eq:hubble_z_de}
\end{align}

So far we did not distinguish neutrinos from the other energy components. After decoupling from the primordial plasma, neutrinos are relativistic and participate to $\rho_r(z)$ along with photons, in a relative amount that scales with the effective number of relativistic neutrinos species $N_{\rm eff}$ ($=3.046$ in the standard model). As Universe expands, the temperature of these relic neutrinos decreases as $a^{-1}$. Due to them being massive, neutrinos become non-relativistic at some redshift $z_{\rm  nr}$; past this point in time, they contribute to $\rho_{\rm m}(z)$. An approximate scaling for a neutrino of mass $m_{\nu}$ reads: $1+z_{\rm nr} \simeq 2000 \times (m_{\nu}/1\,{\rm eV})$. Their contribution to the present-day matter density $\omegam$ writes $\varOmega_{\nu} h^2 = \sum (m_{\nu}/93.14\,{\rm eV})$, with $\sum m_{\nu}$ the (non-zero) sum of neutrino masses. The formula for $\varOmega_{\nu}$ holds even if one of the three neutrino states is still relativistic today \citep{2006LesgourguesPastor}.

The text in this Chapter refers to the usual definitions of distances in cosmology \citep[e.g.][]{1999Hogg}. The angular distance $D_A(z)$ relates the apparent angular size $\delta \theta$ to the proper physical size $D$ of an object located at a redshift $z$, such that $D=\delta \theta D_A(z)$. The luminosity distance $D_L(z)$ relates the bolometric flux $S$ to the bolometric luminosity $L$ of a source at redshift $z$, such that $L = 4 \pi D_L^2(z) S$. Both their expressions involve the evolution factor $E(z)$ introduced in Eq.~\ref{eq:hubble_z} and~\ref{eq:hubble_z_de}.

\subsection{Linear growth of matter perturbations}

The growth of matter inhomogeneities that led to present-day galaxy clusters is described to good approximation in the weak-field limit of General Relativity; in other terms, the Newtonian approximation applies to the evolution of fluctuations on scales $L \ll cH^{-1}$. The derivation focuses on small departures from the homogeneous (`background') model in which structures form.
The Newtonian potential $\Phi$ for an ideal fluid with mass density $\rho$ and pressure $p$ with velocity $v \ll c$ is:
\begin{equation}
    \nabla_{\vec r}^2 \Phi = 4 \pi G (\rho + 3 p/c^2)-\varLambda c^2
\end{equation}
with $\vec r = a(t) \vec x$ the proper separation between two elements in the fluid and $\vec x$ is the comoving separation. The equation of motion is obtained from the geodesic equations and reads $\dd^2 \vec r/ \dd t^2 = -\nabla_{\vec r} \Phi$. Applying this equation to a fixed pair of particles in the background (i.e.~with $\vec x$ held constant) leads to the cosmological background model equation:
\begin{equation}\label{eq:friedmann_equation}
    \frac{1}{a} \frac{\dd^2 a}{\dd t^2} = -\frac{4}{3} \pi G \left[ \rho_b(t) + 3 p_{b}(t)/c^2 \right] + \varLambda c^2/3
\end{equation}
in which index $b$ indicates quantities related to the background homogeneous model. Eq.~\ref{eq:friedmann_equation} is also a consequence of the Einstein field equations and often is called Friedmann's equation. In considering the motion of freely moving particles in the expanding Universe, it is more convenient to form the potential $\phi = \Phi +a \ddot{a} x^2/2$ whose Laplacian with respect to coordinate $\vec x$ is:
\begin{equation}\label{eq:poisson_equation}
    \nabla_{\vec x}^2 \phi = 4 \pi G a^2 \rho_b(t) \delta(\vec x, t)
\end{equation}
with $\delta \equiv [\rho(\vec x, t) / \rho_b(t) -1]$ the overdensity contrast at comoving coordinate $\vec x$. 

The next steps in the derivation involve the description of matter particles as a (ideal, pressureless) fluid in the potential $\phi$, leading to the matter conservation equation and the equation of motion in comoving coordinates. This requires introducing the proper velocity $\vec u = \vec v + \dot{a} \vec x$ of the fluid, with $\vec v = a \dot{\vec x}$ the peculiar velocity of particles relative to the background model.
Combining those equations and linearizing in $\delta \ll 1$ and in $v \ll (\delta L /\tau)$ ($\tau$ being the expansion time of order $\sqrt{G \rho_b}$), one obtains:
\begin{equation}
    \ddot{\delta} + 2 H \dot{\delta} - 4 \pi G \rho_b \delta = 0
\end{equation}
In this equation $H(z)$ acts as a damping term. In particular, as expansion accelerates the damping timescale becomes smaller than the characteristic time for a perturbation to grow and structure growth stops.
The equation accepts a decaying and a growing mode. The latter dominates after some time and the growth factor is $D(t)$ such that:
\begin{equation}
    \delta(\vec x, t) \propto D(t); {\ \ \ } D(z) = \frac{5 \omegam}{2} E(z) \int_z^{\infty} \frac{1+z^{\prime}}{E(z^{\prime})^3} dz^{\prime}
\end{equation}

The steps in this derivation also lead to the evolution of the linerarized peculiar velocity field $\vec v$; it scales linearly with the quantity $f=\dd \ln D/\dd \ln a$, for which a convenient approximation exists:
\begin{alignat}{3}
\label{eq:growth_factor_deriv}
    & f \simeq \omegam(a)^{\gamma} &  \text{\ ;\ with\ } \gamma \simeq 0.55 & & \text{\ and\ }   \omegam(a) = \omegam a^{-3} E(a)^{-2}
\end{alignat}
This approximation is useful in tests of General Relativity probing the evolution of the growth factor.

\subsection{The smoothed linear density field}

Introducing the Fourier transform of the matter overdensity field, $\delta_{\vec k}$, the power spectrum of the linearly evolved matter field at any redshift is such that:
\begin{equation}
    P_L(k, z) \equiv |\delta_{\vec k}|^2 \propto D^2(z) T^2(k) P_0(k)
\end{equation}
Here $P_0$ represents the spectrum of fluctuations at early times, that is typically a scale-free power law of the form $\propto k^{n_s}$, a value $n_s=1$ indicates scale-invariance\footnote{Cosmological microwave background (CMB) measurements provide $n_s = 0.965 \pm 0.004$ \citep{2020PlanckCosmo}.}. The transfer function $T(k)$ quantifies the evolution of each mode relative to the evolution of long-wavelength modes. Its calculation is synthesised in formulas fitted onto numerical simulations. Modern codes \citep[e.g.][]{2000LewisCAMB, 2011LesgourguesCLASS} incorporate in particular the effect of massive neutrinos. Their impact on the power-spectrum -- relative to a spectrum computed with massless neutrinos \citep[e.g.][]{1998EisensteinHu} -- is twofold: i)~a reduction of the growth rate of cold dark matter fluctuations due to the non-vanishing $\varOmega_{\nu}$ term in the expansion rate (Eq.~\ref{eq:hubble_z}), while their contribution to $\rho_b$ in the Poisson equation (Eq~\ref{eq:poisson_equation}) remains negligible at small scales\footnote{This is due to the large free-streaming length of neutrinos preventing their fluctuations to grow on scales smaller than they can travel in a Hubble time $t_H = 1/H$.}; ii)~a shift in the time of matter-radiation equality. A useful approximate formula to gauge the suppression of power in the matter power-spectrum at large $k \gtrsim k_{\rm nr}$ is: $P_L(k, z, f_{\nu}) \simeq -8 f_{\nu} \times P_L(k, z, f_{\nu} =0)$ with $f_{\nu} \equiv \varOmega_{\nu}/\omegam$, as long as $f_{\nu} \lesssim 0.07$ and $z \lesssim z_{\rm nr}$ \citep{1998HuEisensteinTegmark}. The typical scale below which damping occurs is such that $k_{\rm nr} \simeq 0.026 (m_{\nu}/1\,{\rm eV})^{1/2} \omegam^{1/2} h$\,Mpc$^{-1}$.

Smoothing of the linearly evolved density field at a length scale $R$ involves convolution by a spatial filter (or "window function") $W_R(\vec x)$; the variance of the filtered density field writes:
\begin{equation}
    \sigma^2(R, z) \equiv \sigma_R^2 = \frac{1}{2 \pi^2} \int P_L(k, z) |\hat{W_R}(k)|^2 k^2 dk
\end{equation}
where a hat symbol stands for Fourier transform. The common top-hat three-dimensional window function is defined such that it encloses a mass $M=4 \pi R^3 \rho_b/3$ within the sphere of radius $R$. This expression enables defining $\sigma(M, z)$, the variance corresponding to a characteristic mass. The specific value obtained by setting $R=8 h^{-1}$\,Mpc (and $z=0$) defines the $\sigma_8$ parameter which is well constrained by galaxy cluster experiments.

\subsection{Departures from linear growth}

In the hierarchical scenario for structure formation, fluctuations grow linearly until their overdensity contrast reaches $\delta \sim 1$ when they start detaching from the expansion of the background model; their subsequent evolution is driven by gravitational collapse. Relevant notations and concepts are commonly introduced with the simple picture of a spherically symmetric overdensity that is embedded in a uniform background of matter. This spherical collapse model takes roots in early theories of cosmological structure formation.

Spherical collapse equations invoke Birkhoff's theorem in General Relativity in order to describe the evolution of a small perturbation satisfying spherical symmetry in an Einstein-de~Sitter Universe, that is a flat matter-dominated Universe, in which $\rho_b(t) \propto t^{2/3}$ (Eqs.~\ref{eq:hubble_z} and~\ref{eq:friedmann_equation} for $\omegam = 1-\omegak = 1$).
The equations ruling the evolution of the perturbation are similar to those governing the expansion of an overcritical model with perturbed density $\rho \equiv \rho_b (1+\delta)$. Friedmann's equations lead to $\ddot R/R = -GM/R^3$ with $R(t)$ the radius of a sphere enclosing a mass $M=4 \pi \rho_b (1+\delta) R^3/3$. The solution is such that $R$ grows until it reaches a maximum $R_m$ at given time $t_m$ (turn-around time) and then decreases until $t_c = 2 t_m$, time at which the structure collapses ($R \rightarrow 0$). Due to imperfect spherical symmetry and inhomogeneities, complete collapse does not happen. The structure virializes at time $t_{\rm vir} \simeq 2 t_m$ and its radius is fixed at the value given by the Virial theorem, $R_{\rm vir} = R_m/2$, that is reached some time before virialization. Computation of the mass density within $R_{\rm vir}$ provides $\Delta_{\rm vir} = \rho_{\rm vir}/\rho_b(t_{\rm vir}) = 18 \pi^2 \simeq 178$; this value is the typical `virial overdensity' at the time of halo formation in a flat Einstein-de~Sitter Universe.

On the other hand, the evolution of the overdensity contrast $\delta$ may be linearized such that $\delta_{\rm lin} \propto D(t)$. The value of $\delta_{\rm lin}$ extrapolated at $t=t_c$ provides $\delta_c \equiv \delta(t_c) \simeq 1.69$. This characteristic overdensity contrast can be interpreted as a threshold; a structure will collapse if its corresponding linearly extrapolated overdensity reaches $\delta_c$ in the course of its evolution. Therefore, it is convenient to compare the normalized height of peaks $\delta/\sigma(M)$ in the smoothed density field with the threshold $\nu = \delta_c/ \sigma(M)$. Hence, at each redshift $z$ there exists a typical mass $M^*(z)$ describing the scale of collapsing structures; it is such that $\nu(M^*)=1$.

The spherical collapse model is extended to background cosmological models deviating from the flat, matter-dominated Universe. Inclusion of a cosmological constant leads to small corrections on $\delta_c$ \citep{1996KitayamaSuto, 2000Voit}. An evolving equation of state for dark energy translates into small changes in the expression of the spherical collapse threshold \citep[e.g.][]{1998WangSteinhardt, 2003BattyeWeller, 2005HorellouBerge, 2005MaorLahav, 2017Pace}.

A frequent notation for galaxy cluster masses is $M_{\Delta}$. It corresponds to the mass enclosed in a radius $R_{\Delta}$ within which the mean matter overdensity is $\Delta$. A letter indicates whether the overdensity is computed relative to the background $\rho_b$ (with letter `$b$' or `$m$', e.g.~$M_{200b}$, $R_{200m}$) or the critical density $\rho_c$ (e.g.~$M_{500c}$, $R_{500c}$), all densities computed at the redshift of the cluster:
\begin{equation}
\label{eq:mass_delta}
        M_{\Delta c} = \frac{4}{3} \pi R_{\Delta c}^3 \rho_c(z) \Delta \text{\ \ and\ \ } M_{\Delta b}  = \frac{4}{3} \pi R_{\Delta b}^3 \rho_b(z) \Delta
\end{equation}

\subsection{The halo mass function and abundance of clusters}

By identifying collapsed halos of mass $M$ with loci in the smoothed linear density field passing over the threshold $\delta_c$, and with the assumption of Gaussian-distributed $\delta$'s, \citet{1974PressSchechter} derived a decomposition of the halo mass function in the following form:
\begin{equation}\label{eq:massfunction_decomposition}
    \frac{\dd n}{\dd M}(M, z) = f(\sigma) \frac{\rho_b}{M^2} \frac{\dd \ln \sigma^{-1}}{\dd \ln M}
\end{equation}
with $\sigma = \sigma(M, z)$ previously defined and $dn/dM$ the comoving volume density of halos per mass interval. Notably, $f(\sigma)$ presents an exponential cut-off at high masses, with $f(\sigma) \propto \delta_c/\sigma \times \exp\left[-\frac{1}{2}(\delta_c/\sigma)^2 \right]$. This reflects the observational fact that massive clusters are rare entities. Although the original expression for $f(\sigma)$ suffers from several approximations (in addition to those inherent to the spherical collapse model) that makes it too imprecise for current studies, subsequent numerical simulations of structure formation showed the adequacy of this decomposition in reproducing the distribution of halos in standard cosmologies \citep[e.g.][]{1999ShethTormen, 2001Jenkins, 2008Tinker, 2010Pillepich, 2011Courtin, 2016Despali, 2017Comparat, 2019McClintock}. In practice, these studies show that $f$ and its parametrization slightly depend on cosmological parameters and on redshift $z$, at least to precision levels of order $5\%$, that are relevant for precision cosmology.

A more generic formulation for the mass distribution of halos is \citep{2011Courtin}:
\begin{equation}
    \frac{\dd n}{\dd M} = - \frac{\rho_b}{M} \frac{1}{\sigma^2} \frac{\dd \sigma}{\dd M} \int_0^{\infty} \delta \frac{\partial \mathcal{S}}{\partial \delta} \mathcal{F}\left(\frac{\delta}{\sigma}\right)\dd \delta
\end{equation}
where $\mathcal{S}(\delta, \nu)$ is a `selection function' and $\mathcal{F}$ is the probability distribution function of the primordial density fluctuations smoothed on scale $M$. The $\mathcal{S}$-function encapsulates all effects resulting from the non-linear collapse; in the Press-Schechter model that led to the decomposition of Eq.~\ref{eq:massfunction_decomposition}, $\mathcal{S}$ is the Heaviside function on $\delta_c$.

The model presented throughout this Section clarifies the origin of the sensitivity of halo abundances to cosmological parameters. Of them, $\omegam$ and $\sigma_8$ are especially well constrained by local cluster surveys as they strongly influence the number density of massive halos at late times. The parameter $S_8$ is a combination of these two parameters. It is introduced by some authors in an attempt to reduce correlations (i.e.~degeneracy in constraints) while fitting cluster abundance data. Frequently found definitions are $S_8 = \sigma_8 \omegam^{\varGamma}$ and $S_8 = \sigma_8 (\omegam/0.3)^{\varGamma}$, with $\varGamma$ a fixed numerical exponent suited to a specific experiment (typically $\varGamma=0.5$).

%%%%%%%%%%%%%%%%%%%%%%%%%

%%%%%%%%%%%%%%%%%%%%%%%%%%%%%%%%%%%%%%%%%%%%%%
%%%%%%%%%%%%%%%%%%%%%%%%%%%%%%%%%%%%%%%%%%%%%%
\section{Galaxy cluster abundances in X-ray surveys\label{sect:cosmo_abundances}}

Galaxy clusters are rare entities, with present-day volume abundance of the order $10^{-5}$\,Mpc$^{-3}$ at masses above $10^{14}\,M_{\odot}$ and $10^{-8}$\,Mpc$^{-3}$ at masses above $10^{15}\,M_{\odot}$. Determining precisely their abundance is a key to understanding how structure grows in the Universe, what are the processes governing halo formation, and ultimately to constrain cosmological models. Clusters may be counted as a function of any of their properties or combination thereof: flux, luminosity, temperature, etc. Cluster mass is the most interesting property as it is predicted by the halo formation model presented in Sect.~\ref{sect:basic_cosmology}.
At first sight, any count experiment is related to the mass function in some way and all abundance functions are a different representation of this same and unique distribution. However there is no such thing as a direct cluster mass measurement, and comparing various abundance functions enables cross-validations of models, extraction of unexpected systematic biases and tightening of the mass-observable relations. We select and describe recent progress in the counting experiments of galaxy clusters and their derived cosmological constraints.

%%%%%%%%%%%%%%%%%%%%%%%%%%%%
\subsection{X-ray mass estimate: hydrostatic and proxies}

In addition to tracing the hot gas content, the spatial distribution of gas is sensitive to the total mass distribution. The simplest way of solving the dependence is to assume that the gas is in hydrostatic equilibrium (HSE), so that acceleration and velocity terms in the Euler equation can be ignored. This leads to an equation for the encompassed mass within a radius $r$:
\begin{equation}\label{eq:massTX}
M_{\mathrm{HSE}}(<r) = -\dfrac{k_{\mathrm{B}}Tr}{G\mu m_{p}}\left[\dfrac{\mathrm{d}\ln\rho_{\mathrm{gas}}}{\mathrm{d}\ln r} + \dfrac{\mathrm{d}\ln T}{\mathrm{d}\ln r}\right],
\end{equation}

where $T$ is a temperature of the gas and $\rho_{\mathrm{gas}}$ - its density,  $\mu$ - mean molecular weight.
The systematic uncertainty associated with this technique was calibrated using numerical simulations and it is thought to be 10\% \citep{1996Schindler}. These mass estimates were used to derive the first estimate of the gas fraction, and to calibrate the cluster scaling relations \citep{2001Finoguenov, 2006Vikhlinin, 2010Arnaud, 2010Pratt}.

The assumption of gas pressure exactly counterbalancing the gravitational pull is subject to numerous investigations. The reader looking for an entry point into a large body of literature will find in \citet{2019Pratt} a recent review with focus on the importance and the impact of an accurate galaxy cluster mass scale on cosmology.
Numerical simulations play a central role in the examination of the bias and the scatter induced by the HSE assumption. All major simulation results obtained in the last decade indicate an underestimate of $M_{\rm HSE}$ with respect to the true total mass of a galaxy cluster \citep[see e.g.][and references therein]{2021Gianfagna}. The amount of bias depends on the radial range being considered. HSE-based masses within cluster core regions are generally less biased (at the $\sim 5-10\%$ level at $R_{2500}$) than masses within radius $R_{500}$ and beyond, values raising up to $\sim 15-30\%$ at the virial radius \citep{2016Biffi}. The bias depends critically on the dynamical state of the systems and on the mass accretion history of clusters \citep{2012Rasia}. The assumption of HSE is more likely to fail when applied to disturbed systems in contrast to relaxed clusters, and mergers stand out as extreme examples of systems for which a simple HSE hypothesis cannot be applied blindly.
The HSE bias relates to the presence of non-negligible terms in the Euler equation for non-viscous fluids. These terms translates into phenomena familiar to hydrodynamical fluid studies, once given a smoothing scale to perform spatial averaging \citep{2013Lau}: mean gas rotation and streams, random motions, etc. In particular, bulk motions and random turbulent motions are leading the corrections to the HSE assumption and contribute via the acceleration and velocity terms. The expression for random motions may be recast in the form of a component dubbed `non-thermal pressure', adding up to the thermal pressure of the gas $P_{\rm th} \propto \rho_{\rm gas} T$. Simulations indicate an increase of the fraction of non-thermal pressure ranging from $\sim 5-15\%$ at low radii to $20-30\%$ at larger radii, especially in outskirt regions where infall of matter on low-density gas regions prevents from reaching HSE quickly \citep{2014Nelson, 2020Pearce}. Clumping and inhomogeneities in the gas distribution also negatively bias ($\sim 5\%$) the HSE mass, in that they lead to erroneous estimates of the gas density (and temperature) profiles  from X-ray observations, those quantities entering Eq.~\ref{eq:massTX} \citep{2013Roncarelli}.
Simulations indicate moderate values of the  scatter in the $M_{\rm HSE}-M$ relation, of order $\sim 10-15\%$ at low redshift. The predicted evolution of this scatter is not firmly established, some works indicate an increase at earlier cosmological epochs \citep[e.g.][]{2017LeBrun} while other point towards constant spread \citep[e.g.][]{2021Gianfagna}.

On the observational side, results taking as reference weak-lensing-based mass estimates in relaxed clusters within $R_{2500}$ point towards low ($\lesssim 10\%$) bias values. This result is consistent with the numerical simulation estimates once allowance is made for temperature calibration uncertainties \citep[e.g.][]{2016Applegate}. The independent analysis of a low-redshift luminous X-ray cluster sample indicates $10-15\%$ lower X-ray HSE-based masses within $R_{500}$ when compared to weak-lensing estimates \citep{2016Smith}. When considering cluster masses derived from galaxy kinematics as a baseline, a bias value of less than $10\%$ is found at high statistical confidence \citep{2016Maughan}.

In spite of its limitations, the hydrostatic equilibrium equation (Eq.~\ref{eq:massTX}) also illustrates the origin of scaling between the properties of the gas and the total mass, by linking $T\sim \frac{M}{r}$ (for further details and most recent results on the calibration see Chapter ``Scaling relations of clusters and groups, and their evolution''). The presence of this relation is fundamental to use X-ray surveys to count clusters, as parameters of the survey could be linked to the sensitivity towards cluster mass. In addition to cluster detection, parameters of the cluster gas can be used to characterize the cluster mass at lower scatter, compared to e.g. weak lensing mass measurements. These are called mass proxies, with the widely used ones being $L_X$, $T$, $M_{\rm gas}$, and their combination, $Y_X=T M_{\rm gas}$.

%%%%%%%%%%%%%%%%%
\subsection{The X-ray luminosity function}

One of the most studied distribution is the X-ray luminosity function (XLF) of galaxy clusters, $\phi(L_X)$, defined as the number of clusters per unit comoving volume and unit luminosity $L_X$. Constructing such a distribution requires knowledge of cluster redshifts, their fluxes and temperatures (for spectral K-correction) and an estimate of a radius, typically $R_{500}$ that comes from a total mass estimate. Moreover, a precise determination of the volume probed at each luminosity is key and this links directly to the determination of the survey selection function. \citet{2002Rosati} provided a thorough review of \emph{ROSAT}-based luminosity functions; at the time of their review, data was pointing towards mild-to-no evolution of the comoving space density, except that high luminosity clusters were rarer as redshift increases.

\begin{figure}
    \centering
    \includegraphics[width=\linewidth]{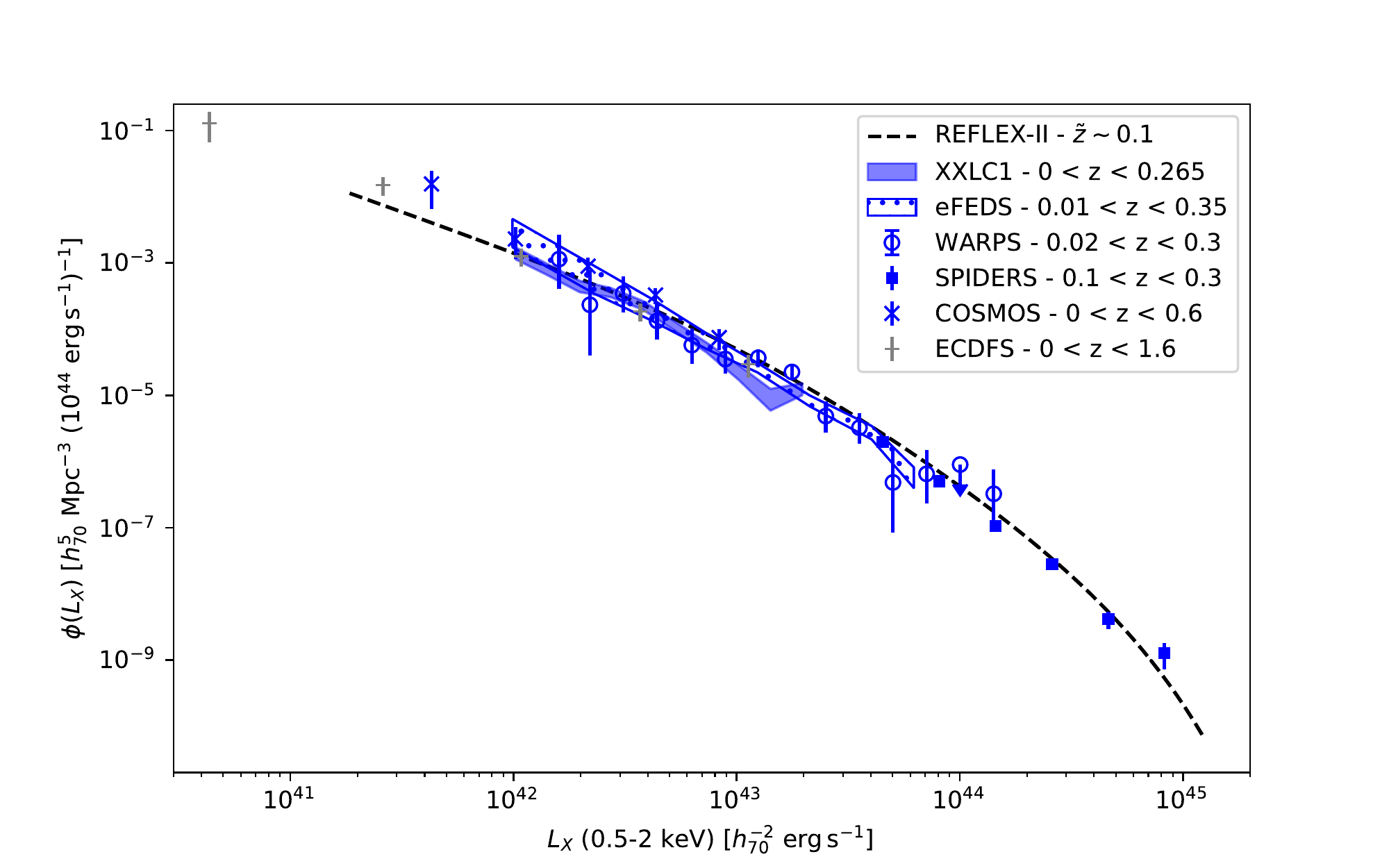}
    \caption{Compilation of low-redshift cluster X-ray luminosity functions from recent surveys. Downward pointed arrows indicate upper limits. References: WARPS \citep{2013Koens}; REFLEX-II \citep{2014Boehringer}; COSMOS+ECDFS \citep{2015Finoguenov}; XXLC1 \citep{2018Adami}; SPIDERS \citep{2020Clerc}; eFEDS \citep{2021LiuEFEDS}.}
    \label{fig:XLF_lowz}
\end{figure}

Figure~\ref{fig:XLF_lowz} displays the luminosity functions in the nearby Universe ($z < 0.3$) established from several new surveys that emerged since then: 160SD \citep{2004Mullis}, WARPS \citep{2013Koens}, REFLEX-II \citep{2014Boehringer}, CDFS \citep{2015Finoguenov}, XMM-XXL \citep{2018Adami}, SPIDERS \citep{2020Clerc} and eFEDS \citep{2021LiuEFEDS}. When needed, the published luminosity values were converted into the 0.5-2 keV rest-frame band. There is good agreement between all measurements across the entire luminosity range. However, deviations occur near a luminosity $\sim 10^{43}$ erg s$^{-1}$. The apparent deficit of clusters in the 50\,deg$^2$ XXL survey is found to be marginally compatible with cosmic variance effects \citep{2016Pacaud}. The WARPS analysis \citep{2013Koens} is pointing to an apparent excess of clusters at around $\sim 2 \times 10^{43}$ erg s$^{-1}$; the comparison with REFLEX-II seems to reduce the significance of this effect.

\begin{figure}
    \centering
    \includegraphics[width=\linewidth]{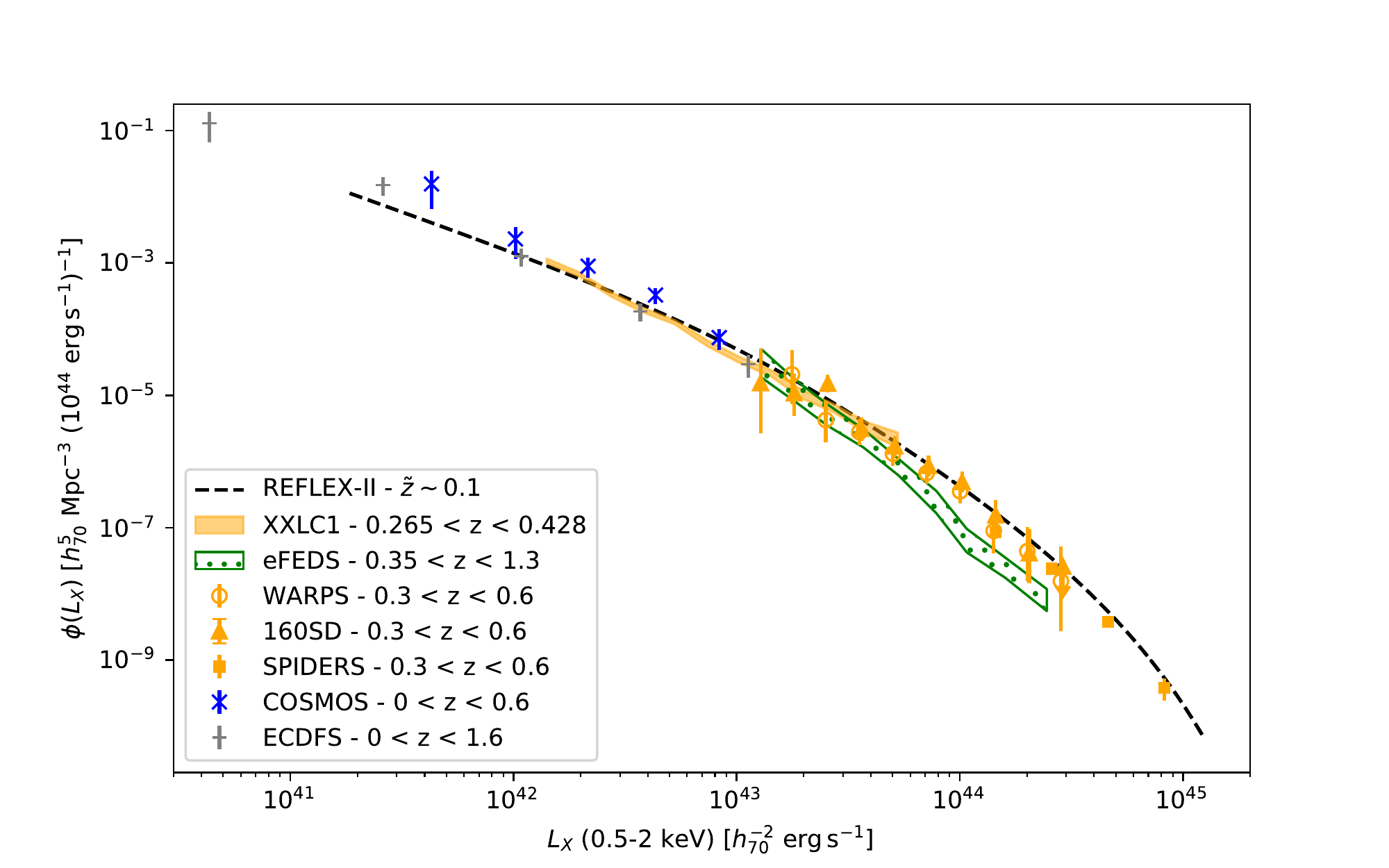}
    \caption{Figure similar to Fig.~\ref{fig:XLF_lowz}, displaying the X-ray luminosity function at higher redshifts, around $z \simeq 0.4$. The local REFLEX-II luminosity function is shown as a dashed line. References: 160SD \citep{2004Mullis} and see Fig.~\ref{fig:XLF_lowz}.}
    \label{fig:XLF_midz}
\end{figure}

\begin{figure}
    \centering
    \includegraphics[width=\linewidth]{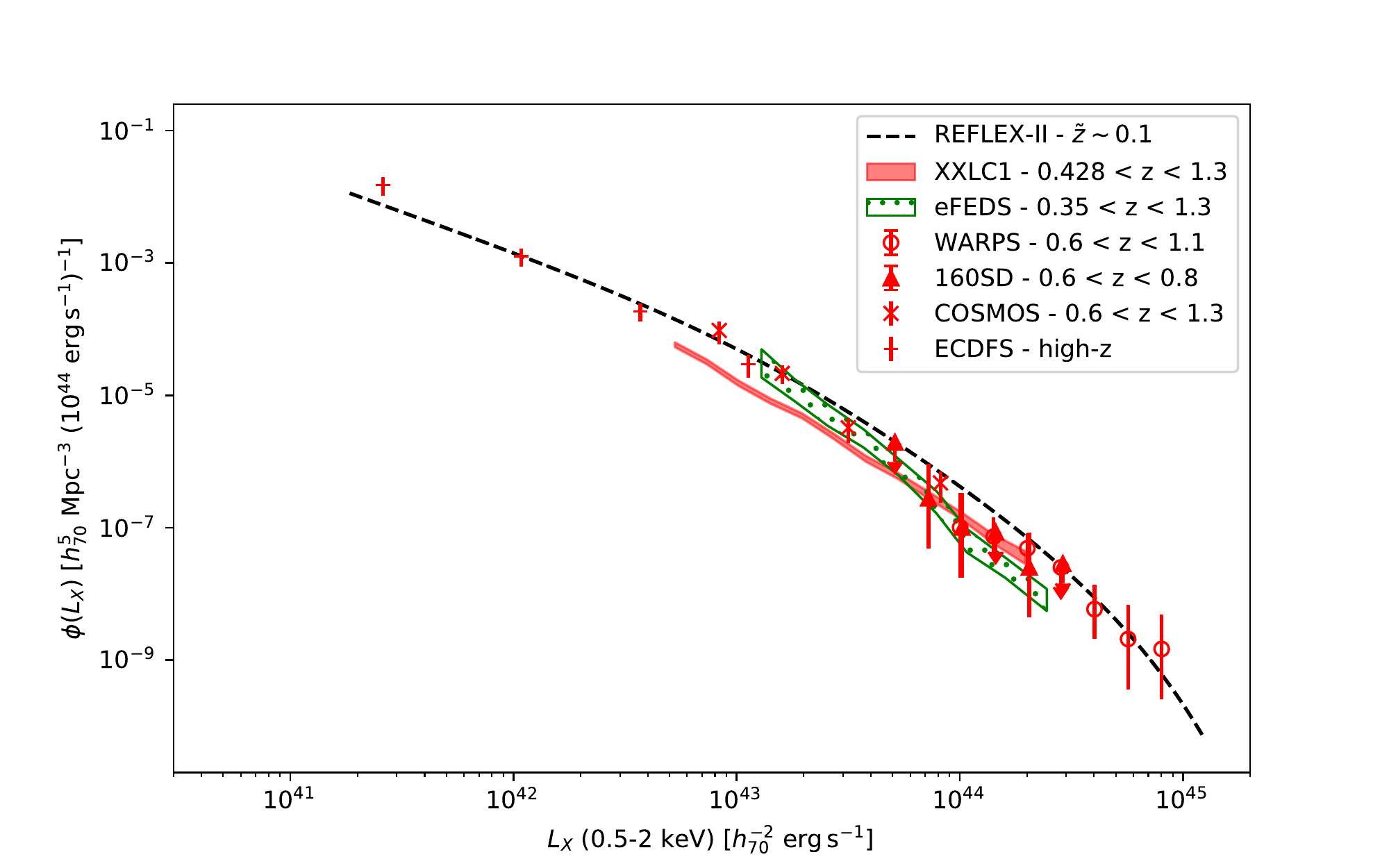}
    \caption{Figure similar to Fig.~\ref{fig:XLF_lowz}, displaying the X-ray luminosity function at higher redshifts, $z > 0.6$. The local REFLEX-II luminosity function is shown as a dashed line. References: see Fig.~\ref{fig:XLF_lowz} and~\ref{fig:XLF_midz}.}
    \label{fig:XLF_highz}
\end{figure}

Figure~\ref{fig:XLF_midz} and ~\ref{fig:XLF_highz} shows the luminosity function computed in redshift shells located above $z=0.3$. This is a major new addition to our understanding of the cluster space density distribution. These high-redshift densities are in agreement with each other, and no evolution of the comoving space density is observed up to $z\sim 0.5-0.6$. Beyond $z \simeq 0.6$, the cluster luminosity functions determined by several surveys point towards a decrease in the normalisation at low-luminosities, while the WARPS analysis \citep{2013Koens} indicates no major evolution at the high-luminosity end.

There are several ways to estimate the XLF from a galaxy cluster sample. A widespread technique involves grouping clusters in bins of luminosity, making easy the visualization of the cluster population and its evolution. It is worth cautioning the reader about biases affecting such `binned' estimators, these biases depending on each survey's completeness limit \citep[e.g.][]{2000PageCarrera}. The XLFs compiled in Figs.~\ref{fig:XLF_lowz}--\ref{fig:XLF_highz} originate from their respective publications and these results rely on various estimators, each has its own bias correction scheme and its own bin-to-bin covariance \citep[see][for a discussion]{2016Pacaud}.

Cosmological constraints from the XLF evolution deduced from bright cluster samples extracted from the RASS led to $\omegam=0.28^{+0.11}_{-0.07}$, $\sigma_8 = 0.78^{+0.11}_{-0.13}$ for a spatially flat model with a cosmological constant and $\omegam=0.24^{+0.15}_{-0.07}$, $\sigma_8=0.85^{+0.13}_{-0.20}$, $w=-1.4 \pm ^{+0.4}_{-0.7}$ \citep{2008Mantz}, see also \citep{2010DelPopolo}. These results were obtained by studying the unbinned (redshift, luminosity) distribution of clusters in the models. These works also discuss the impact of various systematics and prior assumptions on their results, for instance the value of the Hubble constant extracted from CMB observations, and uncertainties on the theoretical mass function. Notably, the transformation from mass to luminosity requires external calibration from local samples \citep[e.g.][]{2002ReiprichBoehringer}. A very similar analysis based on the identical dataset in addition to the 400d cluster sample enables constraining the $\gamma$ parameter entering the definition (see Eq.~\ref{eq:growth_factor_deriv}) of the growth-rate parameter $f(z)$ to $\gamma=0.51_{-0.15}^{+0.16}$ for flat $\varLambda$CDM model \citep{2009Rapetti}. This indicates no departure from General Relativity on cosmological scales.
The XLF of REFLEX-II clusters (dashed line in Fig.~\ref{fig:XLF_lowz}) provided preliminary constraints $\omegam=0.29 \pm 0.04$, $\sigma_8=0.77 \pm 0.08$ \citep{2014Boehringer} in agreement with numerous cluster studies. The reported amount of systematic uncertainties is highlighting the need for more robust scaling relations as the samples grow in size. From the XLF best-fit cosmology and uncertainties, \citet{2017Boehringer} derive a model mass function for clusters in the Universe up to $z \sim 0.25$; in particular this model predicts that 14\% of the total mass in the Universe is found within groups and clusters of masses above $10^{13} M_{\odot}$.

\subsection{The X-ray temperature function}

X-ray spectroscopy provides access to temperature measurements of the intracluster medium. The measurements rely on spectral emissivities predicted by radiative transfer and atomic codes \citep[e.g.][]{1996KaastraSpex, 2001SmithApec} and implemented in widely-used packages such as XSPEC \citep{1996ArnaudXspec} or SPEX \citep{2020KaastraSpexSoftware}. This effort is rewarded by temperature measurements being nearly independent of a cosmological model, hence facilitating the modelling.

Prior to the advent of \emph{XMM-Newton} and \emph{Chandra}, detailed temperature functions (XTF) have been derived, mainly using \emph{ASCA}, \emph{BeppoSAX} and \emph{ROSAT} data, including clusters up to redshift $z \sim 0.8$ \citep{1998Markevitch, 2000Henry, 2001Pierpaoli, 2002Ikebe, 2004Henry}.
Complementing these observations, theoretical studies highlighted the potential complication due to merging systems, for which measurements of the gas temperature and luminosity can be significantly boosted \citep{2002Randall}. In general, the non-homogeneity of the gas density and temperature throughout the intracluster medium has a non-trivial impact on the spectra collected in finite angular apertures. Aided by numerical simulations and assuming an adequate instrumental transfer function, one can relate the measured `spectroscopic-like' temperatures to the simulated gas properties \citep{2004Mazzotta}. 

The exquisite sensitivity, field-of-view and spectral resolution of \emph{Chandra}, \emph{XMM-Newton} and \emph{Suzaku} made possible new measurements of the temperature function. 
Using observations with the \emph{Suzaku} X-ray satellite, \citet{2009ShangScharf} study the local temperature function of a subset of the BCS cluster sample \citep{1998Ebeling}. Following \citet{2002Ikebe}, this work models ICM temperatures with a two-component distribution, the cooler plasma model intends to capture strong cool-core emission\footnote{A sizeable fraction of cluster thermodynamic gas profiles display ubiquitous deviations towards denser, cooler material in their central ($R \lesssim 0.1 R_{500}$) regions. The observational definition for cool-core clusters varies among authors, involving criteria such as central temperature, density, cooling time, entropy. More details are found in the Chapter ``Thermodynamical profiles of clusters and groups, and their evolution'' in this Section of the book.}. The good agreement obtained with previous derivations reflects both the strong sample overlap with earlier studies and the consistency between analyses.
The alternative determination of the temperature function by \citet{2009Henry} accounts for selection effects and sample variance among other systematic effects. The theoretical mass function \citep{2001Jenkins, 2008Tinker} provides a prediction of the temperature distribution of local clusters by means of a temperature-mass relation. This scaling relation is a joint fit of sample data -- hydrostatic and weak-lensing masses being available for a number of objects --, and of hydrodynamic simulation data. Since the sample is X-ray selected, effective area calculation requires knowledge of a luminosity-temperature relation. The values of cosmological parameters obtained in this study are $\sigma_8(\omegam/0.32)^{\alpha} = 0.86 \pm 0.04$ with $\alpha=0.30$  $(0.41)$ for $\omegam \leq 0.32$ $(\omegam \geq 0.32)$.

Instead of relying on a conversion from mass to temperature, some authors proposed a construction of the temperature function of clusters derived from the perturbation of the gravitational potential field \citep[e.g.][]{2015Angrick}; this is perhaps more intuitive, since the gravitational potential is closely related to the temperature of gas particles ($k T/\mu m_p \simeq -\Phi_0/3)$ with $\Phi_0$ being the non-linear gravitational potential depth near the minimum. Such formalism avoids resorting to the mass of a cluster and to the mass-observable relations, both requiring calibration. Such routes are promising in their ability to bypass intermediate, noisy quantities; however it does not remove the need to understand the multi-wavelength selection effects.

In general, temperature measurements require larger integration times than fluxes or luminosities; the signal-to-noise ratios also depend on the temperature itself. For instance, at fixed number of photons ($N \lesssim 10^3$) and with the resolution and sensitivity of \emph{XMM-Newton} EPIC cameras, lower temperature gas is easier to measure thanks to the presence of emission lines, at the expense of the increased uncertainty on metallicity \citep[e.g.][]{2005Willis}. As a result, for reliable temperature measurements of hot ($k T>4$\,keV) clusters, energies below 1 keV are not used \citep[e.g.][]{2010Finoguenov}. Recently, a systematic difference between \emph{XMM-Newton} and eROSITA temperature measurements have been reported \citep{2021Turner}, while luminosity measurements agree. For these reasons, the task of building a temperature distribution for a homogeneous set of galaxy clusters is more difficult than establishing the XLF.

\subsection{The baryon mass function}
As proposed by \citet{2003Vikhlinin}, a measurement of the abundance of clusters as a function of their baryonic mass -- that is a fraction of their total mass -- offers an elegant way to bypass the need for mass-observable scaling relations. Since the baryon fraction in clusters should be close to the universal baryon fraction $f_{b, U} = \omegab/\omegam$, the theoretical halo mass function trivially relates to the baryon mass function through $M_b = \varUpsilon M_{\rm tot} \omegab/\omegam$. The depletion factor $\varUpsilon$ is close to one if clusters are representative of the universal baryon fraction; it is lower than one in general. The masses at fixed baryon overdensity $\delta_b$ must be rescaled by a factor $\varUpsilon^{-(1+\alpha)}$, on top of the $f_{b, U}^{-1}$ factor converting baryonic masses to total masses. The exponent $\alpha$ accounts for the difference between baryon overdensity and total overdensity. It is found $\alpha \simeq 0.5$ for a baryonic overdensity $\delta=324$ (relative to the mean density of the Universe).
Application of this method to a subset of the HIFLUGCS sample \citep{2002ReiprichBoehringer} put constraints on $\omegam h = 0.13 \pm 0.07$ and $\sigma_8 = 0.72 \pm 0.04$ assuming $h=0.71$ \citep{2004Voedvokin}. The degeneracy in constraints between the parameters $(\omegam, \sigma_8)$ slightly differ from the local XTF analyses, at the expense of introducing new parameter dependencies (on $h$, $\omegab h^2$ and $\varUpsilon$).
Studies relying explicitly and solely on the baryon mass function are rare in the recent literature, most of the effort focusing on joint constraints from gas mass fractions and total mass abundances. Caution must be taken when considering a wide range of halo masses: while contribution of baryons locked in stars is small in clusters ($\sim5$\%), it reaches 30\% in low mass groups \citep{2009Giodini, 2012Leauthaud}. In addition, the baryons associated with envelope of central galaxy, which are notoriously hard to measure, play an important role in the total baryon budget \citep{2013Gonzalez, 2018Kravtsov, 2018Furnell}. Moreover, the evolution of the baryon fraction is still uncertain: one expects high-redshift and low-mass systems might lose baryons as a consequence of powerful AGN outbursts.

\subsection{X-ray observable-space distribution: the log$N$-log$S$}
Early studies from the \emph{Einstein} surveys, \emph{ROSAT} (deep and all-sky) surveys and \emph{Chandra} Deep fields enabled the measurement of the abundance of clusters as a function of their apparent X-ray flux $S$ -- usually cast in a cumulative form, the log$N$-log$S$:
\begin{equation}
    N(>S) [\deg^{-2}] = \int_{S}^{+\infty} n(S^{\prime}) \dd S^{\prime} {\ ,  \ \ \ {\rm with}\ } n(S^{\prime}) = \frac{\dd n}{\dd S^{\prime} \dd \Omega}
\end{equation}

It was noted early on that the log$N$-log$S$ distribution derives rather straightforwardly from the XLF, once proper K-correction factors are taken into account (which requires a temperature, redshift and at low temperatures also abundance distribution). Therefore, cosmological information from the cluster mass function is propagated to the log$N$-log$S$, although in a complex form, because of the XLF evolution with redshift, of the mass-luminosity relation and of the selection function \citep[e.g.][]{2008Borgani}.

An observational flux distribution can be derived from survey data once clusters are properly identified among X-ray sources and its construction does not require redshifts. However, the difficulty is to infer the total flux $S$ of the cluster. To do so, one either performs a surface brightness analysis, or one uses the results of deeper observations on a similar mass cluster to link the aperture and total flux. Furthermore one needs to know the radial extent to which $S$ is measured (e.g. 1\,arcmin in XXL, $R_{500c}$ in CODEX \citep{2020Finoguenov}, MCXC \citep{2011Piffaretti}, deep surveys) and the level of point source contamination.
A statistical correction for the area $\Omega(S)$ probed at each X-ray flux (i.e.~the selection function) is needed to account for missing systems in the reconstruction of the true log$N$-log$S$.
For a purely flux-limited catalogue the correction is independent on the underlying cluster population. For instance, in a homogeneous survey with geometrical area $\Omega$, one would write: $\Omega(S) = \Omega \times \varTheta(S-S_{\lim})$ with $\varTheta$ the Heaviside function. This formulation assumes that $S \equiv S^{obs}$ is the measured flux (dependent on the observation, signal-to-noise ratio, etc.), in contrast to the true flux $S^{true}$ of a cluster.

In theory, accounting for flux measurement errors $P(S^{obs} | S^{true})$ and making an assumption on true underlying log$N$-log$S^{true}$ of clusters enables to properly correct for sample incompleteness:
\begin{equation}
    \frac{\dd n}{\dd S^{obs} \dd\Omega} = \frac{\Omega(S^{obs})}{\Omega} \int_{0}^{+\infty} \dd S^{true} \frac{\dd n}{\dd S^{true} \dd \Omega} P(S^{obs} | S^{true})
\end{equation}
However, the interesting quantity here is $dn/dS^{true}$ and inversion of this equation is not necessarily a straightforward task. One may resort to forward-modelling techniques \citep{2020Finoguenov, 2020Comparat}; deconvolution is also an option.
In practice, such complications are often neglected and one may compute an effective $\Omega^{\rm eff}(S^{true})$ through simulations of realistic observations. These simulations incorporate a known cluster flux distribution, and the effective area $\Omega^{\rm eff}$ is obtained with the ratio of observed versus true number of sources. The same simulations provide a mapping in the form : $P(S^{true} | S^{obs})$. An approximate method to recover the log$N$-log$S^{true}$ from an observed sample then consists in calculating the log$N$-log$S$ in terms of the observed flux, then resampling true flux values around each $S^{obs}$ and finally dividing the result by $\Omega^{\rm eff}$. If measurement uncertainties are negligible and the function linking observed and true flux is invertible, a simple change of variables suffices.

Another complication arises since more realistic selection functions $P(I | S^{obs}, \mathbf{\theta})$ call for a model capturing additional cluster population properties $\mathbf{\theta}$, in a form that reads:
\begin{equation}
    \label{eq:observable_distribution}
    \frac{\dd n}{\dd S^{obs} \dd\Omega} = \int \dd \mathbf{\theta} P(I | S^{obs}, \mathbf{\theta} ) P(S^{obs} | \mathbf{\theta} ) \frac{\dd n}{\dd \mathbf{\theta} \dd \Omega}
\end{equation}
In these formulas, the notation $P(I| \text{parameters})$ represents the probability of including a source in a sample given a set of parameters.
The vector represented by $\mathbf{\theta}$ may contain random variables $\mathbf{\theta_d}$  that play an explicit role in the detection process (e.g. cluster shapes or number of photon hits deposited on the detectors) and random variables $\mathbf{\theta_m}$ that do not explicitly enter the detection chain. Writing $\mathbf{\theta} = (\mathbf{\theta_d}, \mathbf{\theta_m})$, we have:
\begin{equation}
    \label{eq:observable_distribution_model}
    \frac{\dd n}{\dd S^{obs} \dd\Omega} = \int \dd \mathbf{\theta_d} P(I | S^{obs}, \mathbf{\theta_d}) \int \dd \mathbf{\theta_m} P(S^{obs}, \mathbf{\theta_d} | \mathbf{\theta_m}) \frac{\dd n}{\dd \mathbf{\theta_m} \dd\Omega}
\end{equation}

Notably this expression does not involve the quantity we are looking for, that is the underlying true flux distribution. It may be obtained by calculating $n(S^{true}) = \int \dd \mathbf{\theta} P(S^{true} | \mathbf{\theta}) n(\mathbf{\theta})$, once the distribution $n(\mathbf{\theta})$ is determined from Eq.~\ref{eq:observable_distribution} or~\ref{eq:observable_distribution_model}.

This short discussion sheds light on a difficulty encountered in comparing log$N$-log$S$ derived from different surveys, each using different methods to approximate the required corrections. Let us consider for instance a selection depending only on the number of cluster-emitted photons hitting the detector in a survey, i.e.~$\mathbf{\theta_d} \equiv \eta$. The correlation between $S^{obs}$ and $\mathbf{\theta_d}$ may vanish in an experiment where fluxes come from independent follow-up observations; on the contrary, it equals one if flux is computed directly from the survey counts (e.g.~if $S^{obs} \propto \eta$). With this modification only, the observed flux histograms from these two experiments slightly differ because of the two different joint distributions $P(S^{obs}, \mathbf{\theta_d} | \mathbf{\theta_m})$ that enter Eq.~\ref{eq:observable_distribution_model}.
Fortunately, flux measurement uncertainties and biases are often negligible in practice and selection functions depend mainly on $S^{obs}$ at first order, making the above formalism somehow too demanding for standard usage.

\subsection{X-ray observable-space distribution: general observables}

The formalism exposed in the context of the log$N$-log$S$ may be generalized to quantities other than flux. Most recent cosmological constraints from cluster abundances rely on a similar hierarchical Bayesian structure.

Taking advantage of the spectral capabilities of \emph{XMM-Newton}, flux can be measured in energy bands of $\Delta E \lesssim 1$\,keV in order to form hardness ratios\footnote{A standard definition for hardness ratios is $HR=H/S$ with $H$ the X-ray count-rate (or flux) in a `hard' band, $S$ in a `soft' band. For details refer to Chapter~``X-ray color analysis'' in this book.}. These bands should be narrow enough to capture broad spectral variations and still sufficiently large to preserve signal and to remain insensitive to small-scale spectral features. In absence of any other information, hardness ratios carry information about the temperature and redshift of clusters \citep[e.g.][]{2003BartelmannWhite}.
By modelling the distribution of clusters in the \emph{XMM-Newton} count-rate/hardness ratio space -- i.e. replacing $S^{obs}$ by $(CR, HR)$ in Eq.~\ref{eq:observable_distribution} --, \citet{2012Clerc} found that cosmological constraints are obtainable from a sizeable sample of clusters with moderate individual signal-to-noise data.
The descriptive power of the observable-space cluster distribution increases with increasing number of observables, or dimensions. The process of adding redshift information to the flux distribution of clusters is closely related to studying the XLF evolution; this was a technique employed to constrain $\omegam$ and $\sigma_8$ from the RDCS sample \citep{2001Borgani}. Adding in the distribution of cluster apparent sizes as measured on X-ray images and an extra hardness ratio enables tighter constraints on cosmological and scaling relation parameters \citep{2017Pierre, 2018Valotti}. The advantage of such methods lies in their ability to provide population models as close as possible to measurements, those measurements being devoid of any model assumption. One downside is the potentially infinite number of observable parameters to choose from and the high computational demand in modelling multi-dimensional distributions.

\citet{2020Grandis} develop a formalism applying to multi-wavelength observables and selections. A cluster number count analysis combining RASS (X-ray) and DES\footnote{The Dark Energy Survey covers 5000\,deg$^2$ in the Southern sky in five optical bands.} (optical) relies on a likelihood function involving the density of clusters per bin of X-ray flux ($S$) and optical richness ($\lambda$). This density may be rewritten in the form of Eq.~\ref{eq:observable_distribution_model} with $\mathbf{\theta_d} \equiv (S^{obs}, \lambda^{obs})$ and $\mathbf{\theta_m} \equiv (S^{true}, \lambda^{true})$.

A framework is proposed by \citet{2019Mantz} in the context of performing regression on truncated data. It is illustrated with examples specific to modelling galaxy cluster scaling relations, when some systems are absent from a selected sample. It shares similarities with the formalism presented in this Chapter (in particular Eq.~\ref{eq:observable_distribution_model}); these works underline the tight coupling between two apparently distinct tasks, one of constraining the cosmological parameters and the other of deriving the cluster scaling relations.

\subsection{Recent cluster abundance studies}

\citet{2009Vikhlinin} present cosmological parameter constraints obtained from \emph{Chandra} observations of 37 clusters with $\langle z\rangle = 0.55$ derived from 400 deg$^2$ \emph{ROSAT} serendipitous survey and 49 brightest $z \approx 0.05$ clusters detected in the All-Sky Survey. Evolution of the mass function between these redshifts requires $\varOmega_\Lambda > 0$ with a $\sim 5\sigma$ significance, and constrains the dark energy equation-of-state parameter to $w_0 = -1.14 \pm 0.21$, assuming a constant $w$ and a flat universe. Fitting their cluster data jointly with the latest supernovae, Wilkinson Microwave Anisotropy Probe, and baryonic acoustic oscillation measurements, they obtain $w_0 = -0.991 \pm 0.045$ (stat) $\pm 0.039$ (sys). The joint analysis of these four data sets puts a conservative upper limit on the masses of light neutrinos $\sum m_\nu < 0.33$\,eV at 95\% CL. 

The study of the abundance of massive and bright RASS clusters by \citet{2010Mantz} improves on previous analyses by resorting to deep X-ray follow-up of a substantial fraction of the sample. Their statistical framework distinguishes between observed quantities and intrinsic quantities and accounts for selection effects and scaling relations in a self-consistent manner. The addition of 50 weak-lensing mass measurements (more exactly: shear profiles) among the 224 clusters brings new information on the absolute mass scale of clusters \citep{2015Mantz}. Although weak-lensing masses are individually noisy, they are thought to be of little bias and their inclusion in the model leads to stronger constraints on virtually all model parameters -- although the scatter in individual weak-lensing masses prevents from a large increase in precision on parameters sensitive to redshift evolution, such as $w$. Adding in the (reportedly independent) gas fraction measurements of relaxed clusters in the radial range $0.8 < r/R_{2500} < 1.2$ by \citet{2014Mantz}, the analysis leads to $\omegam = 0.26 \pm 0.03$ and $\sigma_8 = 0.83 \pm 0.04$. The study of \citet{2015Mantz} explores various extensions to the flat $\varLambda$CDM model, in particular it constrains $\omegal$ to $0.73 \pm 0.12$ with a strong preference for zero metric curvature. The equation of state of dark energy is also constrained with good precision ($w = -0.99 \pm 0.06$) as well as evolving equation of state ($w_0=-1.04^{+0.13}_{-0.18}$, $w_1=0.3^{+0.4}_{-0.6}$). Interestingly, the adopted mass calibration does not provide a low value of $\sigma_8$, in contrast to other contemporaneous cluster studies; this translates into a constraint on $\sum m_{\nu}$ that is compatible with zero.

With the help of \emph{Chandra} data, \citet{2017SchellenbergerReiprich} revised the analysis of the HIFLUGCS sample \citep{2002ReiprichBoehringer} consisting of 64 galaxy clusters and groups located at $z \lesssim 0.1$ and primarily selected upon their high apparent X-ray flux in RASS. Measurements of hydrostatic masses and gas masses for all clusters provide the basis of the gas fraction test (or $f_{\rm gas}$, see Sect.~\ref{sect:standard_candles}). This test relies on the well-determined value of $\omegab$ from CMB data, and on the assumption that baryon cluster content is close to representative of the whole Universe baryon content. The relationship between the two is parameterized as a function of mass, redshift and radius with priors from numerical simulations. This test delivers a posterior confidence range for $\omegam$; the mass function analysis in turn may take this as a prior in the cosmology fit. The slope and normalisation of the luminosity-mass relation are fit simultaneously with cosmological parameters. The best-fit values are $\omegam = 0.22^{+0.07}_{-0.05}$ and $\sigma_8=0.89 \pm 0.010$ (without $f_{\rm gas}$ priors on $\omegam$) and $\omegam = 0.30 \pm 0.01$ and $\sigma_8 = 0.79 \pm 0.03$ (with $f_{\rm gas}$ prior). This study describes a number of systematics and corrections; among those the extrapolation of mass, gas temperature and density profiles for truncated observations and the impact of baryons on theoretical mass function. In particular, the inclusion of groups and dynamically disturbed systems is found to bias the fit towards low values of $\omegam$. This bias is due either to sample incompleteness at low mass, or to a different luminosity-mass relation in this regime. Such a modification in the scaling relation possibly originates from the impact of baryonic physics onto observable quantities that differs between clusters and groups.

Focusing on a lower-mass regime, the high-significance subset of 178 galaxy groups and clusters in the 50\,$\deg^2$ XXL survey extends out to $z \sim 1$. The distribution of these systems along the redshift dimension is modelled in \citet{2018Pacaud} and leads to the following constraints: $\omegam = 0.316 \pm 0.060$, $\sigma_8 = 0.814 \pm 0.054$ and $w = -1.02 \pm 0.20$. Despite the very different ranges of masses probed by both surveys, the $\varLambda$CDM fit (fixing $w=-1$) agrees well with \emph{Planck}-SZ cosmology. Given the uncertainties it does not show any significant tension with \emph{Planck}-CMB cosmology. The relation between mass and observables -- necessary to model the selection of objects -- involves scaling relations derived from the same sample (or a subset thereof). In particular the mass-temperature relation relies on weak-lensing mass measurements.

CODEX \citep{2020Finoguenov} show an analysis of the XLF in a $\varLambda$CDM context using 24,788 RASS sources identified in the $0.1<z<0.6$ range as galaxy clusters with redMaPPer5.2 run on 10,382 square degrees of SDSS\footnote{The Sloan Digital Sky Survey has created a wide map of the Northern sky in five optical bands.} photometry. Under assumption of self-similar evolution of scaling laws, and a 5\% error achieved on the $L_X-M$ calibration,  and taking into account the survey selection both at X-rays and in the optical, they find $\omegam = 0.270 \pm 0.06 \pm 0.015$ and $\sigma_8 = 0.79 \pm 0.05 \pm 0.015$. 

A subset of the CODEX clusters confirmed with optical spectroscopic follow-up \citep{2021Kirkpatrick} led to precise ($\Delta z \sim 0.001$) cluster redshifts. These 691 sources benefit from optical richness measurements based on DESI\footnote{The Dark Energy Spectroscopic Instrument will obtain optical spectra for tens of millions of galaxies and quasars.} preparation imaging data that serve as a mass proxy in the cosmological analysis of \citet{2020IderChitham}. There, X-rays enter only in the selection of the systems; an additional redshift-dependent richness selection helps in increasing the purity of the sample and in raising the spectroscopic confirmation rate. The resulting constraints read $\omegam= 0.34^{+0.09}_{-0.05}$ and $\sigma_8 = 0.73 \pm 0.03$. Replacing richness by 1-dimensional galaxy velocity dispersions as a mass proxy, these read $\omegam= 0.33 \pm 0.02$ and $\sigma_8 = 0.74^{+0.03}_{-0.02}$ \citep{2021Kirkpatrick}. In the latter analysis, a theoretical scaling between the total mass and observed velocity dispersion is used, which decreases the uncertainties.

In comparison of different results, it is important to separate the measurements that constrain the same process, where results should be the same, with extrapolations (for instance, cluster abundance studies are directly comparable to the large-scale-structure cosmology using weak lensing shear and galaxy correlation function). Recent results by KiDS\footnote{The Kilo-Degree Survey has mapped 1350\,deg$^2$ of the sky in four broad-band optical filters.} and DES confirm the somewhat lower values of $S_8$ inferred from cluster studies. As an example of possible solutions to a tension with \emph{Planck}-CMB cosmology, \citet{2015BoehringerChon} showed that a non-zero neutrino mass may reconcile the REFLEX-II galaxy cluster XLF best-fit $\sigma_8$ with the (2014) \emph{Planck}-CMB best-fit amplitude of the matter fluctuation power-spectrum; although uncertainties in the mass-luminosity relation of clusters prevents from getting precise constraints on the neutrino masses. Although the required value of the neutrino mass is not supported by the final \emph{Planck}-CMB results, consistent values were reported using a combination of ACT and SPT and WMAP\footnote{The Atacama Cosmology Telescope, the South Pole Telescope and the Wilkinson Microwave Anisotropy Probe provide maps of CMB anisotropies, either from ground or from space.} \citep{2021DiValentino}. Thus, this possibility remains open for the future studies, e.g. with new CMB experiments. 

As illustrated by \citet{2020Eckert}, X-ray emission is a low-scatter mass proxy at $\sim 0.5 R_{500c}$. The scatter increases towards $R_{500c}$ and it is not well known beyond it. \citet{2019Mulroy} showed that the scatter in Compton $Y$ parameter depends on the resolution of the measurement, as inclusion of the zone outside of $R_{500c}$ in \emph{Planck} measurements, leads to larger scatter. X-ray morphology studies have been used to separate out regular clusters. Cluster detection based on the concentration of red sequence galaxies using comparable radii is only possible for rich end of the cluster mass function \citep{2019Costanzi}. Spectroscopic selection of galaxy groups often uses $R_{90c}$, in order to improve the sensitivity, which leads to a large variety of dynamical states of the detected objects, not all of which are containing massive virialized halo.

%%%%%%%%%%%%%%%%%%%%%%%%%%%%%%%%%%%%%%%%%%%%%%
%%%%%%%%%%%%%%%%%%%%%%%%%%%%%%%%%%%%%%%%%%%%%%

\section{Clusters as tracers of large-scale structure\label{sect:clusters_clustering}}

It is well-recognized that collapsed halos cluster in a different manner from the underlying matter field \citep[e.g.][]{1984Kaiser}. A simplified understanding may be gained by considering the superposition of two random fields fluctuating on much different spatial scales. The slowly varying random field $S$ is supposed to have smaller amplitude. In terms of typical variances, $\sigma_S^2 < \sigma_N^2$; hence $S$ `modulates' the rapidly evolving field $N$. In absence of modulation, objects form at locations where $\delta_N/\sigma_N > \nu$ and the probability for an object to form, $P_N(>\nu)$, is given by the properties of the noise field. In presence of the field $S$, the resulting overdensity field is $\delta = \delta_S + \delta_N$ and the probability for an object to form at location $\vec x$ is spatially modulated; it writes:
\[
P(>\nu, \vec x) \simeq P_N(> \nu-\delta_S/\sigma_N) \simeq P_N(>\nu) \left[1+B \delta_S(\vec x)\right]
\]
The second equality is a first-order Taylor expansion with $B$ being expressed in terms of $P_N$, its derivative, and $\sigma$. Hence, the overdensity contrast of objects at location $\vec x$ relative to the average density of objects writes $\delta_h(\vec x) = \left[P(>\nu, \vec x)/P_N(>\nu)\right] - 1 \simeq B \delta_S(\vec x)$. The ratio between the object and `background' overdensity fields, $\delta_h/\delta_S = B$ translates into a 'bias' factor $B^2$ in the ratio of spatial correlation functions of objects and of the field $S$, as is clear from the definition of the correlation function $\xi$: 
\[
\xi_{hh}(r) = \langle \delta_h(\vec x) \delta_h(\vec x + \vec r) \rangle \simeq B^2 \langle \delta_S(\vec x) \delta_S(\vec x + \vec r) \rangle = B^2 \xi_{SS}(r)
\]

\subsection{Two-point clustering of halos and the bias parameter}

Beyond this illustrative picture, the definition for bias is well established in the case of random fields representative of the large-scale matter distribution \citep[e.g.][]{1986Bardeen}. On spatial scales large enough to be insensitive to the non-linear evolution of the density field, the ratio between the halo and matter power-spectra defines the bias parameter $b^2$. The `peak-background split formalism' provides an useful means to estimate the variation of bias with object mass \citep[e.g.][]{1986Bardeen, 1996MoWhite, 1999ShethTormen}; it involves knowledge of the mass function of halos conditional on the background density and a prescription for relating Lagrangian to Eulerian overdensities \citep[e.g.][]{2007Zentner}. Calibration of the large-scale halo bias using numerical simulations allows capturing departures from the peak-background split model and accurate fitting formulas provide state-of-the-art values for the bias. These are usually given as a function of mass, $b(M)$, or as a function of peak height, $b(\nu)$, for several mass definitions and overdensity thresholds. Two such formulations for the large-scale halo bias are $b_{\rm T10}$ \citep{2010Tinker} and $b_{\rm B11}$ \citep{2011Bhattacharya}, expressed as:
\begin{align}
    b_{\rm T10}(\nu) &  = 1 - A \frac{\nu^a}{\nu^a + \delta_c^a} + B \nu^b + C \nu^c \ \ \ \label{eq:bias_tinker10}\\
    b_{\rm B11}(\nu) & = 1 + \frac{a \nu^2-q}{\delta_c} + \frac{2 p/\delta_c}{1+\left( a \nu^2 \right)^p} \label{eq:bias_comparat17}
\end{align}
In both formulas, $\nu = \delta_c/\sigma(M,z)$ is related to a halo mass $M_{\Delta}$. Eq.~\ref{eq:bias_tinker10} is calibrated on N-body simulations with $(a, b, c, A, B, C)$ being dependent on the overdensity $\Delta$ and independent on redshift \citep{2010Tinker}. Eq.~\ref{eq:bias_comparat17} derives from the peak-background split formalism applied to the halo mass functions established in \citet{2011Bhattacharya}; the fitting parameters $(a, p, q)$ therefore intervene in the expression for $f(\sigma)$ (as defined by Eq.~\ref{eq:massfunction_decomposition}) and may \citep{2011Bhattacharya} or may not \citep{2017Comparat} depend on redshift.

\subsection{Constraints from X-ray clusters two-point clustering analyses}

Due to the sensitivity of the clustering amplitude of dark matter halos of a given mass to the underlying cosmology the application of the clustering theory to galaxy clusters is theoretically highly motivated. Moreover, galaxy clusters exhibit scaling relations between their baryonic properties and the total mass of their hosting halos. These properties include among other the cluster X-ray luminosity and richness (that is, number of galaxies belonging to the cluster). Using on one hand these observable quantities as mass proxies and on the other hand the connection between cluster masses and their bias one can make inference on cosmological model. Most attempts to follow this route have, however, resulted in cosmological parameters, which are in discordance with the constraints obtained using the number counts of the same sample \citep{2002Schuecker} or lead to strongly disagreeing scaling relations \citep{2017Jimeno}. Remarkably, \cite{2012Allevato} obtain a precise agreement between the modelling of the clustering, based on the weak lensing mass calibration and the measured clustering amplitudes of galaxy groups, by taking a detailed account on the definition of the object and the effects of sample variance.

The power-spectrum analysis of the REFLEX-II sample \citep{2014Boehringer} is presented in \citet{2011BalagueraAntolinez} with a comparison and validation supported by large mock catalogues mimicking the original survey selection. The X-ray luminosity-dependent clustering bias of clusters is modelled both in redshift- and real-space by means of a mass-luminosity relation calibrated such as to recover the REFLEX-II XLF; the effective bias is an average weighted over the mass distribution of clusters. Covariance matrices are obtained from the mock catalogues, highlighting the strength of off-diagonal terms due to mode-coupling, especially at small scales. The systematic distortions of the power-spectrum induced by the flux-limited nature of the sample are potentially important limitation in the analysis of the REFLEX-II sample, akin to Malmquist bias in abundance studies \citep[e.g.][]{2001Schuecker}. The analysis of the mocks indicate negligible effect in the range of wave numbers of interest. The amplitude of the power-spectrum derived from subsamples selected upon varying luminosity thresholds shows good agreement with predictions from the $\varLambda$CDM model. While Baryon Acoustic Oscillations (BAO) signatures are not expected to be distinguishable given the obtained signal-to-noise ratios, the shape of the power-spectrum hints for departures from the linear structure growth at intermediate scales.

Despite its relatively small sky coverage, the increased sensitivity of the  50\,deg$^2$-wide XXL survey probes low cluster masses (median $M_{500} \simeq 10^{14} M_{\odot}$) at high redshift (median $\bar z \simeq 0.3$). The clustering analysis of 182 high-significance X-ray-detected systems in both the Northern and Southern halves of the survey \citep{2018Marulli} constrains $\omegam$ and the effective bias of the cluster sample $b_{\rm eff}$, to $0.27^{+0.06}_{-0.04}$ and $2.73^{+0.18}_{-0.20}$ respectively. The latter constraint matches well the bias predicted in the $\varLambda$CDM model, the mass of XXL clusters being derived from the weak-lensing $M_{wl}-T_X$ relation as measured from a subset of those systems. Limited statistics prevent from fitting additional parameters, as well as marginalizing over uncertainties in the mass-observable relations. The constraints from the XXL survey ultimately should follow from a combination of abundance and clustering analyses \citep{2011Pierre}.

\begin{figure}
    \centering
    \includegraphics[width=\linewidth]{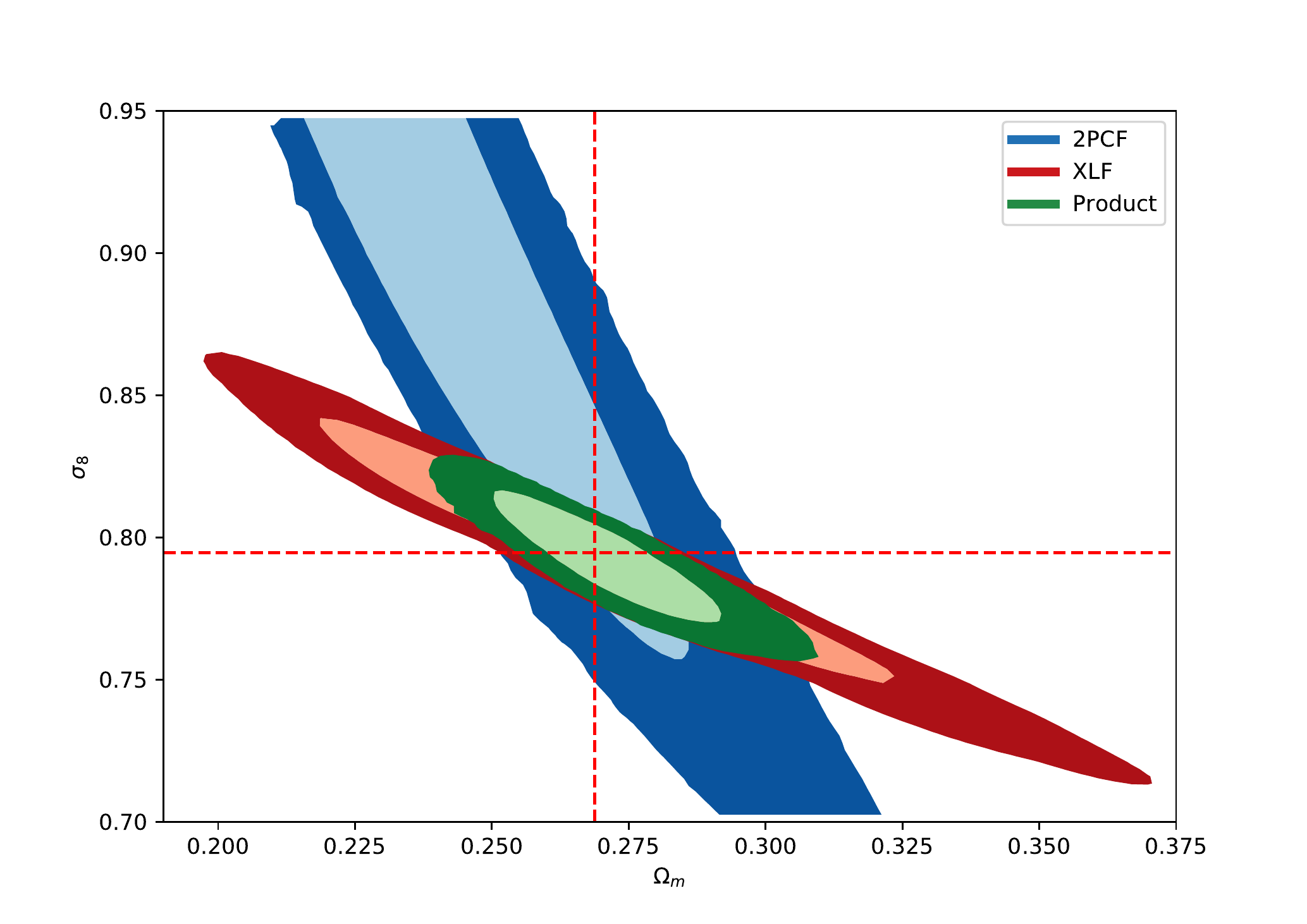}
    \caption{The likelihood function of $(\omegam$, $\sigma_8)$ from the cluster two-point correlation function (blue), X-ray luminosity function (red) and joint distribution (green). The light and dark contours are the 68 \% and 95 \% confidence regions, respectively and red dashed lines show the best fit values for the joint likelihood. Credit: Lindholm et al., A\&A 646, A8, 2021, reproduced with permission \copyright ESO.}
    \label{fig:lindholm_2pcfxxlf}
\end{figure}

The analysis of the CODEX galaxy clusters \citep{2021Lindholm} aims at predicting the clustering bias of these systems based on their masses. All three components in the analysis -- the clustering of galaxy clusters (or dark matter halos in the case of simulations), the expected dark matter distribution and the mass-to-bias conversion -- depend on the assumed cosmology. Bringing them into agreement therefore provides a test of the cosmological model.
A parallel analysis of the halo catalog built from a (4 $h^{-1}$ Gpc)$^3$ dark matter simulation \citep{2016Klypin} shows the ability of this approach to recover $\omegam$ and $\sigma_8$ within statistical uncertainties. The likelihood involves a comparison between the two-point correlation function of halos and the total matter power-spectrum rescaled by the bias factor, as predicted from the mass of the systems. CODEX cluster masses are estimated either from their X-ray luminosities or their optical richnesses; these estimates predict bias factors in agreement with the measured value at the level of 17-39\%, with richness-based estimates providing better agreement. Splitting the sample in redshifts bins leads to predicted bias values agreeing with the measured values at the 5-17\% level, depending on the redshift range.
The clustering analysis in redshift bins provides the following constraints: $\omegam=0.22^{+0.04}_{-0.03}$ and $\sigma_8=0.98^{+0.19}_{-0.15}$. Additional systematic uncertainties of $\pm 0.02$ and $\pm 0.19$ respectively, arise due to survey effects and bias modelling. Combining the clustering-based likelihood with one from the cluster mass function the following constraints follow, also represented in Fig.~\ref{fig:lindholm_2pcfxxlf}: $\omegam = 0.27^{+0.01}_{-0.02}  {\rm (stat.)} \pm 0.07 {\rm (syst.)}$ and $\sigma_8 = 0.79^{+0.02}_{-0.02} {\rm (stat.)} \pm 0.04  {\rm (syst.)}$. The consideration of covariance errors makes an estimate of their contribution below the statistical error of this analysis, while this results shows a clear dominance of systematics, within which no difference to \emph{Planck}-CMB cosmology could be claimed. 

Forecasts for large cluster survey experiments combine the abundance and clustering into a powerful probe of cosmological models. Of specific interest there are departures from Gaussian initial fluctuations\footnote{\label{note:planckNG}Although recent results from \emph{Planck}-CMB rule out high levels of departure from Gaussian initial perturbations \citep{2020PlanckCMBNG}.}. A popular, albeit not unique, parametrization of primordial non-Gaussianities involves the dimensionless parameter $f_{\rm NL}$ quantifying the amount of quadratic corrections added to the Gaussian potential; it relates to the skewness in the distribution of the overdensity contrast $\delta$. In a work addressing X-ray surveys, \citet{2010Sartoris} predict constraints on primordial non-Gaussianity involving both the power-spectrum and number counts from samples of sizes $\gtrsim 10^5$. This study demonstrates their complementary strengths, particularly in breaking degeneracy between $f_{\rm NL}$ and $\sigma_8$. The specific case of the eROSITA all-sky survey, promising $\mathcal{O}(10^5)$ clusters, is studied in \citet{2018Pillepich}. Several scenarios are investigated; the most favourable one involves X-ray follow-up observations of a subset of objects in order to constrain the luminosity-mass relation at a level beyond that currently known. In this case, the expected one-sigma uncertainty on $\omegam$, $\sigma_8$, $w_0$ and $w_1$ amount to $\pm 0.006$, $\pm 0.008$, $\pm 0.07$ and $\pm 0.25$ respectively, after combination with complementary cosmological probes (CMB, BAO, Type Ia Supernovas).

%%%%%%%%%%%%%%%%%%%%%%%%%%%%%%%%%%%%%%%%%%%%%%

\section{Sample variance considerations}

%Computation of the volume-driven error on the mass function of galaxy clusters.

Sample variance originates from observations of finite volume of the Universe and the fact that large-scale structure is spatially correlated. Counts or clustering observables extracted from surveys of any size are affected by this additional term, regardless of the mass of the tracers \citep[e.g.][]{1996MoWhite}.

    \subsection{Variance in cluster number counts}

The relative sample covariance $\sigma_{ij}^2$ between two estimates of galaxy cluster number density $n_i(M_i, z_i)$ measured in two volumes described by their window functions $W_i(\vec x)$ such that $\int \dd r \dd\Omega W_i  = 1$ may write \citep{2003HuKravtsov}:
\begin{equation}
\label{eq:hu_kravtsov}
    \sigma_{i,j}^2 = \frac{\langle n_i n_j \rangle-\bar n_i \bar n_j}{\bar n_i \bar n_j} = b(M_i, z_i) b(M_j, z_j) D(z_i) D(z_j) \int \frac{\dd^2 \vec k}{(2 \pi)^3} {\hat W_i}(\vec k) {\hat W_j}^*(\vec k) P(k)
\end{equation}
where $\bar n$ is the average density of clusters, $P(k)$ is the linear matter power-spectrum today, $D(z)$ is the linear growth-rate as a function of redshift, $b(M,z)$ is the linear bias parameter for objects of mass $M$ at redshift $z$ and three-dimensional Fourier transforms are annotated with a hat. The additional variance term due to cluster counts following a Poisson distribution is usually denominated `shot-noise variance'; it adds quadratically to sample covariance.
\citet{2003HuKravtsov} consider the rather general case of a survey window function that is separable into an angular mask and a selection varying with the radial direction. For number counts in two radial bins indexed by $i$ and $j$ in the local Universe, Eq.~\ref{eq:hu_kravtsov} reads:
\begin{equation}
    \label{eq:hu_kravtsov_spherical}
    \frac{\langle n_i n_j \rangle - \bar n^2}{\bar n^2} = b^2 \sum_{l, m} 4\pi \int \frac{\dd k}{k} {\tilde R}_{il}(k) {\tilde R}_{jl}(k) \left| {\tilde \Theta}_{lm} \right|^2 \frac{k^3 P(k)}{2 \pi^2}
\end{equation}
where window functions are such that $W_i = R_i(r) \Theta(\theta, \phi)$. ${\tilde R}_{il}(k)$ is the Bessel transform of the radial window, ${\tilde \Theta}_{lm}$ is the harmonic transform of the angular mask. Similar equations found successful application to galaxy surveys \citep[e.g.][]{2002Newman, 2008Trenti, 2011Moster}.

For local all-sky cluster surveys comprising a single mass and redshift bin and extending out to a radius $R$, the variance term reduces to multiplying $b^2$ by $\sigma_R^2$, the variance of the linear matter field smoothed over a radius $R$. The scaling of sample variance with the survey volume $V \propto R^3$ depends on the shape of $P(k)$. For nearly flat power-spectra, sample variance thus scales as $1/V$. Shot-noise relative variance scales as $1/\bar n V$, therefore for such survey configurations, the relative importance of sample variance and shot-noise depends only weakly on survey volume. For a given local survey volume, the relative importance of shot-noise increases with increasing mass selection threshold, due to the rarity of high-mass clusters.

\begin{figure}
    \centering
    \includegraphics[width=\linewidth]{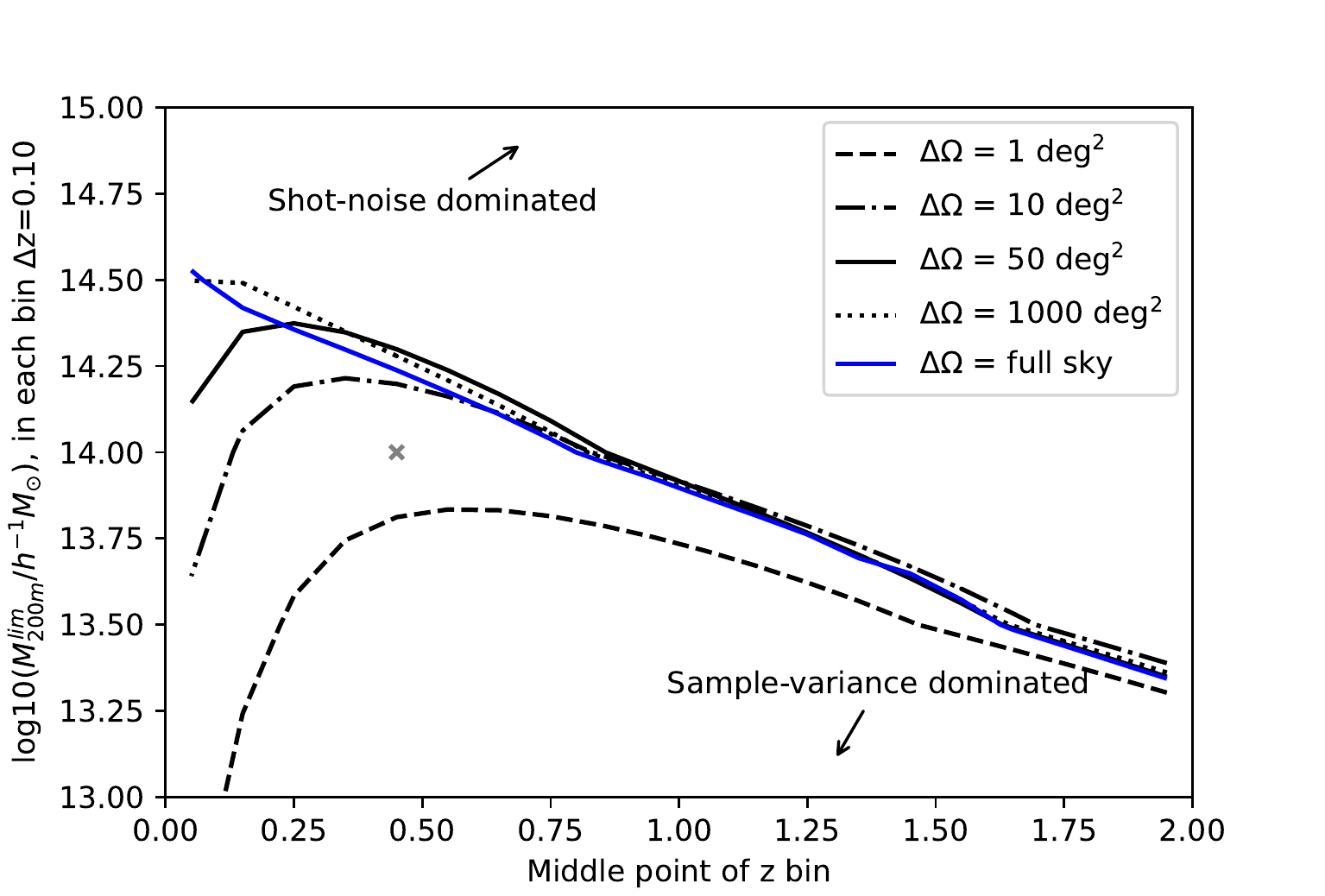}
    \caption{Estimating the relative amount of sample variance and shot-noise in galaxy cluster count experiments in broad redshift bins. Any point in this diagram corresponds to a limiting mass $M_{200}^{\rm lim}$ at given redshift $z$. The curves delineate the boundary between regions in this diagram where sky number densities in $\Delta z = 0.1$ bins are shot-noise dominated (above) and sample-variance dominated (below). Each curve is calculated following the equations in \citet{2011Valageas} and assuming a `spherical cap' survey geometry covering a solid angle $\Delta \Omega$. As an example, the variance in the number density of $M > 10^{14} h^{-1} M_{\odot}$ clusters satisfying $0.4<z<0.5$ \emph{(grey cross)} is dominated by shot-noise in a $1 \deg^2$ survey; but it is dominated by sample-variance in a full-sky survey.}
    \label{fig:sn_equals_sv_area}
\end{figure}

An example use of the formalism is provided in Fig.~\ref{fig:sn_equals_sv_area} assuming the \emph{Planck}-CMB cosmology \citep{2016PlanckCosmo}. Focusing on the number $\mathcal{N}(>M^{\rm lim})$ of clusters per solid angle in $\Delta z = 0.1$ bins above a certain mass threshold, the diagram displays the loci of survey configurations for which the uncertainty on $\mathcal{N}$ is dominated by shot-noise or by sample variance. For a given redshift bin, deep surveys corresponding to low threshold masses provide large amounts of clusters; for those surveys sample-variance considerations are relatively important. The boundary between the two regimes depends on the area covered -- here a spherical cap geometry was assumed in our calculation. The location of these boundaries additionally depends on the binning in redshift and on the upper mass selection threshold.

Formulas such as Eqs.~\ref{eq:hu_kravtsov},~\ref{eq:hu_kravtsov_spherical} may be implemented with the help of a library such as COLOSSUS \citep{2018Diemer}. As an application, we find the following empirical fit for the ratio between the sample variance and shot-noise variance in the sky number density $\mathcal{N}$ (units deg$^{-2}$) of clusters above a given mass $M_{200m}^{\rm lim}$ in a narrow bin of redshift $\Delta z \lesssim 0.1$:

\begin{equation}
       \frac{\sigma_{\rm sv}^2}{\sigma_{\rm sn}^2} \approx \frac{n(>M^{\rm lim}, z)}{1.5 \times 10^{-6} \left(h^{-1} \mathrm{Mpc}\right)^{-3}} \left[ \frac{\left\langle b(z) \right\rangle_{> M^{\rm lim}}}{4}\right]^2 \times 0.4^{(1+z)} 
\end{equation}

This crude approximation holds for a continuous survey coverage of area greater than $1000$\,deg$^2$, at redshifts $0.1 \lesssim z \lesssim 2$ and $10^{13} M_{\odot} \lesssim M_{200m}^{\rm lim} \lesssim$\,few\,$10^{14} M_{\odot}$ in a \emph{Planck}-CMB cosmology. The halo bias $b(z)$ is averaged over the population of clusters satisfying $M > M^{\rm lim}$ and the volume density of halos $n$ has units $(h^{-1} \mathrm{Mpc})^{-3}$.

Accounting for a mass- and redshift-dependent cluster selection function in Eq.~\ref{eq:hu_kravtsov_spherical} involves considering the volume probed by the survey at a given mass index; the radial bins then consist of overlapping, increasingly large spheres. \citet{2003HuKravtsov} examined the specific case of a local, flux-limited X-ray survey with $R = (1+z)^{-1} \sqrt{L_X(M)/(4\pi f_{lim})}$, with $L_X(M)$ the scatter-less mass-luminosity relation.
In practice, X-ray cluster selection is not similar to a mass threshold and the derivation of sample variance should be reassessed with the exact selection cuts.
For surveys extending further out in redshift, the relative importance of shot-noise variance and sample variance depends on the sky coverage, the mass selection threshold and the maximal redshift. Shot-noise is found to dominate at high redshift where massive halos are rare.
For the specific cases of XMM-XXL and the eROSITA all-sky survey the transition redshifts between the sample-variance-dominated and shot-noise-dominated number count experiments in $\Delta z \sim 0.1$ bins are $z\sim 0.7$ and $z\sim 0.5$ respectively.

    \subsection{Extensions of the sample variance formalism}

The analytic formalism of \cite{2003HuKravtsov} is extended beyond number counts by \citet{2011Valageas} to the sample variance of the halo correlation function $\xi(r)$, for various estimators including the Landy-Szalay one; the effects of redshift-space distortions discussed in \citet{2012Valageas} are of tiny impact.

\citet{2013TakadaHu} dubbed 'super-sample covariance' the effects due to sampling the power-spectrum of the matter in a finite Universe volume. In their proposed approximation, the modes with inverse wave numbers exceeding the survey size act in a similar fashion as a uniform rescaling of the density background. Calibration of the power-spectrum response to a perturbation $\delta_b$ of background density provides a term additive to the power-spectrum covariance. In the frame of the halo model, halo sample variance appears as a limiting case of super-sample covariance in the strongly non-linear regime. % -- that is the regime suited to cluster counts analyses.
A concise and generic formulation for 'super-sample covariance' between two observables $\mathcal{O}_1$ and $\mathcal{O}_2$ collected in two redshift bins is \citep[e.g.][]{2016LacasaRosenfeld}:
\begin{equation}
    \label{eq:lacasa_SSC}
    {\rm Cov}_{\rm SSC} \left(\mathcal{O}_1, \mathcal{O}_2\right) = \int \dd V_1 \dd V_2 \frac{\partial o_1}{\partial \delta_b}(z_1) \frac{\partial o_2}{\partial \delta_b}(z_2) \sigma^2_{\rm proj}(z_1, z_2)
\end{equation}
where the integral runs over two volumes defined by the redshift bins and the angular window functions. The term $\sigma^2(z_1, z_2)$ is the covariance of the background density fluctuation $\langle \delta_b(z_1) \delta_b(z_2) \rangle$, related to $P(k)$ and to the growth rate $D(z)$. The comoving density of the observable, $o_i$, reacts to a change in background density as $\partial o_i/\partial \delta_b$. Specifically for cluster number counts experiments, $\mathcal{O}_1 \equiv N(i_M, i_z)$ and $\mathcal{O}_2 \equiv N(j_M, j_z)$ are the numbers of clusters in two mass and redshift bins indexed by $(i_M, i_z)$ and  $(j_M, j_z)$. Thus $o_1 = n(i_M, z)$ and $o_2 = n(j_M, z)$ are the densities of clusters in those mass bins, and the following holds:
\begin{equation}
     \frac{\partial n}{\partial \delta_b}(i_M, z) = \int_{M \in i_M} \dd M \frac{\dd n}{\dd M} b(M, z)
\end{equation}
which is the average of the linear halo bias weighted by the halo mass function $dn/dM$.
Following this formalism, \citet{2018Lacasa} demonstrate that for large redshift bins (of size $\Delta z \simeq 0.1$), the formula in Eq.~\ref{eq:hu_kravtsov} underestimates the sample variance by $\sim30\%$. In fact, the derivation in \citet{2003HuKravtsov} implicitly assumes non-evolving bias within a window function support domain. Both descriptions (Eq.~\ref{eq:hu_kravtsov} and~\ref{eq:lacasa_SSC}) therefore coincide in the limit of small redshift bins. Correlations between redshift bins that are large compared to the angular size of the survey can be ignored and this leads to a computationally simpler expression \citep{2011Valageas, 2017KrauseEifler}.

Since galaxy cluster bias $b$ is a prediction from the cosmological model, the variance in cluster abundance across multiple patches of the sky encloses constraining power. With simple assumptions on the sample selection mechanism and assuming a low (known) scatter in the mass-observable relation, \citet{2004LimaHu} show how full accounting for sample variance effects can break degeneracies induced by a mere number count analysis. Their likelihood represents a joint fit of cosmological parameters and of the mass-observable relation. Information on the evolution of the scatter and bias can be retrieved at the expense of additional binning of the data with a mass proxy \citep{2005LimaHu}.
Covariance among count-in-cells also probes primordial non-Gaussianity of the local type thanks to the strong scale dependence of the halo bias and the coupling of cells on scales of hundreds of Mpc \citep{2009Oguri, 2010Cunha}.

Sample variance is also a concern for numerical simulations aiming at reproducing the surveys under study. \citet{2019KlypinPrada} estimate that multiple simulations boxes with sizes $\sim(1 - 1.5) h^{-1}$\,Gpc are suited to most cosmological studies, including galaxy cluster abundance studies, without significant loss of power. Running such a large volume with full hydro-dynamical simulation represents a substantial investment of computing resources.

%%%%%%%%%%%%%%%%%%%%%%%%%%%%%%%%%%%%%%%%%%%%%%
%%%%%%%%%%%%%%%%%%%%%%%%%%%%%%%%%%%%%%%%%%%%%%

\section{Clusters as standard candles\label{sect:standard_candles}}

Cosmological constraints often rely on standardizable probes to perform cosmography or evolutionary experiments -- a typical example being Type Ia supernovae and their use as standard candles. Galaxy clusters enter this class of cosmological studies in complementing the tests relying on population distributions described previously.

\subsection{The gas fraction tests}
In Sect.~\ref{sect:cosmo_abundances} we presented a cosmological test based on the abundance of clusters as a function of their baryonic mass. Similar arguments provide support to a quite different and powerful test that consists in measuring the gas mass fraction ($f_{\rm gas}=M_{\rm gas}/M_{\rm tot}$, the ratio of gas and total masses) in galaxy clusters, and in relating this value to the Universal $f_{b, U} = \omegab/\omegam$ prediction. Unlike abundance studies, truncation of samples due to selection effects is less of a concern; the main difficulty consists in choosing systems for which only small departures from spherical symmetry and hydrostatic equilibrium are expected. They should be massive enough for their matter content to be representative of the whole Universe and for having a reduced stellar fraction. Under such conditions the universal baryon fraction should approximately match $f_{\rm gas}$ with a small amount of cluster-to-cluster variance.
With robust priors on $\omegab h^2$ and the Hubble constant, the application of this test at low redshift provides tight constraints on $\omegam$ \citep[see e.g.][for an early application]{1993White}. Furthermore, assuming a limited evolution of $f_{\rm gas}$ with time (as indicated by numerical simulations incorporating cooling and AGN feedback) serves in constraining parameters governing the evolution of the luminosity and angular diameter distances, hence the cosmological parameters entering $H(z)$.
Subsequent to the results and analyses of \citet{2008Allen} and \citet{2009Ettori} putting constraints on $\omegam$, $\omegal$ and $w$, \citet{2014Mantz} presented a new sample of gas fraction measurements in 40 massive and hot clusters extending out to redshift $z \sim 1.1$. A noticeable feature is the systematic selection of dynamically relaxed clusters based on their X-ray appearance in deep \emph{Chandra} images. Gas fraction measurements in a radial range $[0.8, 1.2] R_{2500}$ constitute the observable that is compared to the following model \citep{2014Mantz}:
\begin{equation}
    \label{eq:fgas_mantz}
    f_{\rm gas}^{\rm ref}\left( 0.8 < r/R_{2500}^{\rm ref} < 1.2 \right) = K(z) A(z) \varUpsilon(z) \left(\frac{\omegab}{\omegam}\right) \left[ \frac{d^{\rm ref}(z)}{d(z)}\right]^{3/2}
\end{equation}
where $d(z)$ is either the luminosity ($D_L$) or angular ($D_A$) diameter distance, $K=K_0 (1+K_1 z)$ represents the ratio between total and X-ray mass, modelled with a linear dependence with redshift. $A$ accounts for the shift in radial range with changing cosmology, $\varUpsilon =  \varUpsilon_0 (1+\varUpsilon_1 z)$ is the gas depletion factor scaling linearly with redshift. The superscripts ``ref'' indicate the values for the reference cosmology fixed in the process of measuring $f_{\rm gas}$. In the \citet{2014Mantz} analysis, lensing masses for 12 clusters provide constraints on $K_0$, as all cluster masses are derived after deprojection of temperature and density profiles under the assumption of hydrostatic equilibrium. Numerical simulations \citep{2013Planelles, 2013Battaglia} give the prior on $\varUpsilon_0$. The analysis assumes uniform priors on $K_1$ and $\varUpsilon_1$ centered around zero.
The low-redshift data leads to tight constraints on $h^{3/2} \omegab/\omegam$. Once combined with local Hubble constant measurements and Big Bang Nucleosynthesis, these translate into $\omegam = 0.27 \pm 0.04$. For non-flat $\varLambda$CDM models, the full sample provides $\omegal = 0.65^{+0.17}_{-0.22}$ and $\omegam= 0.29 \pm 0.04$. Models considering the equation of state of dark energy in flat $\varLambda$CDM are constrained with good accuracy thanks to the redshift leverage arm in the sample and a reduced amount of systematic uncertainties in the derivation.

Modifications or extensions to the model above (Eq.~\ref{eq:fgas_mantz}) can be exploited to test alternative cosmological scenarios.
The analysis of \citet{2017Magana} combines $f_{\rm gas}$ samples established in X-ray analyses \citep{2008Allen} and in SZ  \citep{2013Hasselfield} to examine various models for the equation of state of Dark Energy. Some of these models tend to marginally indicate a slowing down of cosmic acceleration at late times, although this result vanishes as CMB, BAO and Hubble constant measurements are included in the analysis.
\citet{2016LiHeGao} argue that $f(R)$ theories of gravitation induce a change in the relation between gas temperature (hence, the inferred X-ray hydrostatic mass) and the true cluster mass in comparison to standard gravitational theory. This change should lead to a modification of the relation between $f_{\rm gas}$ and $\omegab/\omegam$; their analysis of the \citet{2008Allen} sample thus excludes strong departures from general relativity.
Focusing on departures that break the duality relation between luminosity and angular diameter distances (see below), \citet{2017Holanda} use the sample of $f_{\rm gas}$ by \citet{2014Mantz} in combination with temperature measurements of the CMB at various redshifts in order to establish consistency with general relativity.

\subsection{Distance measurements with combined X-ray and SZ observations}
The inverse Compton scattering of photons constituting the cosmological background radiation (CMB) by energetic electrons in the intracluster medium distorts the blackbody CMB spectrum. This process called the Sunyaev-Zeldovich (SZ) effect imprints characteristic $\Delta T/T_{\rm CMB} \sim 10^{-5}$ secondary temperature fluctuations in the CMB sky observed at frequencies around $\nu \simeq 100-300$\,GHz, typically over arcminute angular scales corresponding to a cluster apparent extent \citep{1972SunyaevZeldovich}.
In the Rayleigh-Jeans part of the CMB spectrum, the temperature decrement imputed to the thermal SZ effect reads:
\begin{equation}
    \Delta T_{\rm CMB} \simeq -2 \frac{k T_{\rm CMB}}{m_e c^2} \int \sigma_{T} n_e T_e \dd l
\end{equation}
with $\sigma_T$ the Thomson cross-section, $T_{e}$ the electron temperature and the integral is performed along the line of sight. The dependence on $n_e$ and $T_e$ in the integrand differs from the X-ray surface brightness dependence (Eq.~\ref{eq:xray_surface_brightness}).

A cluster of typical size $\mathcal{L}$ along the line of sight imprints a SZ decrement scaling as $\Delta T/T \sim \mathcal{L} \overline{n_e T_e}$ (a bar indicates averaging along the line of sight). On the other hand its X-ray surface brightness scales as $S_X \sim \mathcal{L} \overline{n_e^2 \Lambda_{\rm cool}}$, with $\Lambda_{\rm cool}$ the X-ray cooling function, dependent mainly on gas temperature and elemental abundance. Assuming spherical symmetry of the intracluster gas, $\mathcal{L}$ can be obtained from a measurement of the cluster angular extent $\Delta \theta$; cancelling $n_e$ in both expressions thus provides an estimate of the angular diameter distance at the redshift of the cluster, $D_A(z)=\mathcal{L}/\Delta \theta$, that can be compared with the prediction of the cosmological model.

The argument presented here in a simplified form supported early attempts at combining X-ray and millimeter-wavelength observations of galaxy clusters, with the goal of retrieving the Hubble constant $H_0$ and the deceleration parameter $q_0=-\left.\ddot a a/\dot a^2 \right|_{t=0}$ \citep[e.g.][]{1978SilkWhite, 1979Cavaliere}.
The assumption of spherical symmetry may be dropped by performing sample averages and the test can incorporate more sophisticated gas shapes. Consequently this test is also sensitive to the distribution of cluster shapes in a sample; by this means \citet{2005DeFilippis} discovered a bias in a sample of 25 X-ray selected clusters being preferentially elongated and aligned along the line of sight.
Under the assumption of spherical symmetry and hydrostatic equilibrium of the gas, \emph{Chandra} X-ray data and resolved observations of the SZ-effect in 38 clusters at redshifts $0.1 < z < 0.9$ lead to the value of $H_0=76.9^{+3.9}_{-3.4}$\,(stat.)\,$^{+10.1}_{-8.2}$\,(syst.)\,km\,s$^{-1}$\,Mpc$^{-1}$ for a $\varLambda$CDM cosmology with $\omegam=0.3$ and $\omegal=0.7$ \citep{2006Bonamente}.

The expression for the X-ray surface brightness can be rewritten with an explicit dependence on the luminosity and the angular diameter distances:
\begin{equation}\label{eq:xray_surface_brightness}
    S_X = \frac{\eta^2(z)}{4\pi (1+z)^4} \int n_e^2 \Lambda_{\rm cool} \dd l
\end{equation}
with $\eta(z) = D_A(z)(1+z)^2/D_L(z)$. It makes apparent the result of the X-ray/SZ test being a measurement of $D_A/\eta^2$ rather than $D_A$. It thus provides a check of the `distance-duality relation', that is an observational test of the equation $\eta=1$ predicted by the standard cosmological model \citep{2004Uzan}. To this end, an estimate of the distance luminosity at the redshift of the clusters is necessary. Assuming WMAP-1 cosmology, the sample of X/SZ distance measurements of \citet{2002Reese} shows no significant violation of the $\eta=1$ relation \citep{2004Uzan}. This results holds if instead of assuming a cosmological model, supernova distance moduli measurements provide $D_L(z)$ in a model-independent fashion \citep[e.g.][]{2012Holanda, 2013Liang}.

\subsection{Recent results on the Hubble constant measurements}

The combined analysis of \emph{Planck} SZ-effect measurements and \emph{XMM-Newton} data for 61 clusters at $z<0.5$ leads to $H_0=67 \pm 3$\,km\,s$^{-1}$\,Mpc$^{-1}$ \citep{2019Kozmanyan}. The analysis takes advantage of resolved SZ and X-ray data to jointly model the pressure, density and temperature profiles in clusters. The ratio between the pressure profiles derived from each observable is decomposed in two terms. The first term involves cosmology dependencies, among them $H_0$ and the helium abundance; the second term incorporates what is described in the next paragraph as systematic uncertainties, i.e.~clumpiness, asphericity, etc. The latter component is constrained via the use of hydrodynamic simulations \citep{2015Rasia} providing a quantitative prior on its distribution.

An independent analysis of the pressure profiles of 14 dynamically relaxed clusters \citep[selected following the criteria in][]{2014Mantz} combined with \emph{Planck} and Bolocam S-Z data provides $H_0=67.3^{+21.3}_{-13.3}$\,km\,s$^{-1}$\,Mpc$^{-1}$ \citep{2021WanMantz}. The apparent larger uncertainties originate from inclusion of extra freedom in the analysis, aiming at characterizing the (currently uncertain) temperature measurement bias specific to the \emph{Chandra} calibration. This analysis demonstrates such a calibration uncertainty is the major source of error in their determination of $H_0$. Reversing the argument and assuming $H_0$ is known from CMB data or from Cepheid variables, the sample is used to constrain the temperature measurement bias.

\subsection{Sources of systematic uncertainties}

The $f_{\rm gas}$ tests resort to galaxy cluster total mass measurements. Cosmological results are therefore sensitive to the biases and uncertainties in those measurements, such as the hydrostatic bias introduced in the discussion of Eq.~\ref{eq:massTX}. Minimizing the bias and scatter around true mass values may be achieved by carefully selecting morphologically relaxed and isolated clusters, those systems being more likely to display low mass accretion rate and gentle dynamical disturbance. Calibration of measured masses against numerical simulations provide expectations and prior values for the deviations from truth.

Clumping of gas on scales smaller than the resolution of the instrument leads to an enhancement of the X-ray emission, leading typically to a positive bias in the value of $H_0$ inferred from the X/SZ test.
\citet{2010Finoguenov} looked at the source of clumping suggesting that pressure fluctuations contribute only 5\% and the main source is entropy fluctuations. The latter can be removed in X-ray images using the choice of the special band from imaging \citep{2016Churazov}.
Systematic uncertainties in the galactic absorbing column density and the absolute calibration of the temperature measurement are small, but relevant effects to consider when using clusters as standard candles \citep[e.g.][]{2006Bonamente}.

An additional source of uncertainty stems from the gradient in helium (He) abundance within the cluster. Numerical simulations indicate moderate helium sedimentation in the low-temperature outskirts of clusters, however core regions are potentially richer in helium due to the drift of higher-mass atoms toward cluster centres \citep[e.g.][]{2009PengNagai}. Not accounting for such gradients may induce biases in the recovered gas densities in central regions, due to X-ray emissivities being function of the abundances of all atoms, including He \citep[e.g.][]{2000QinWu, 2011Bulbul}. This translates into a systematic uncertainty on the measurement of $H_0$, $f_{\rm gas}$, total masses and thus on cosmological parameters. As an example, \citet{2009PengNagai} predict up to 10\% bias on $H_0$ from the X/SZ test performed within $R_{2500}$ if helium sedimentation is unaccounted for. Reversing the argument, prior knowledge on the Hubble constant and cluster shapes enables constraining the fraction of Helium in central cluster regions \citep{2007Markevitch}.

\subsection{Distance measurements from spectra of X-ray resonant lines}

The X/SZ test presented above involves measurements of the cluster size along the line-of-sight, specifically a quantity proportional to $\mathcal{L} \overline{n_e}$ that is the SZ Compton-$y$ parameter. A similar observable can be constructed in the X-ray domain, considering the optical depth $\tau \propto \mathcal{L} \overline{n_e}$ of resonant lines in high-resolution spectra of X-ray light passing through the intracluster medium.
While most lines in the ICM X-ray spectrum are optically thin $(\tau \ll 1)$, some strong lines show optical thicknesses $\tau \gtrsim 1$, such as \ion{Fe}{XXV} at 6.7\,keV in big clusters. Incoming light at the line frequency is absorbed and reemitted in a random direction. Resonant scattering thus modifies the shape, decreases the amplitude \citep[e.g.][]{1980Shapiro, 1987Gilfanov, 2010Churazov} and possibly imprints polarization \citep[e.g.][]{2002Sazonov} of these lines and at the same time it leaves intact non-resonant lines.

Measurements of the optical depth may involve resonant absorption lines in the spectra of light passing through the galaxy cluster and emitted either from a background AGN \citep{1988Krolik} or -- most likely -- from an AGN hosted in a cluster core \citep{1989Sarazin}. This technique involves comparisons of on-target (AGN+ICM) observations with off-target (ICM only) emission line measurements to estimate the optical thickness and, in turn, the angular diameter distance to these clusters and constraints on the Hubble constant and its evolution.

Alternatively, resonant emission lines originating from the ICM itself may be used in measuring the optical depth. \citet{2006Molnar} estimate a few thousand photons in the ICM iron 6.7~keV resonant line at $2-4$\,eV spectral resolution lead to a 5\%-error measurement on $\tau$, provided $\tau \gtrsim 2$; their calculation predicts a 10\% statistical uncertainty in the measurement of $D_A(z)$ from 100 clusters at $0<z<1$, which translates into a 16\% uncertainty (3-$\sigma$ level) on $\omegam$ in a flat Universe.
In both cases (absorption or emission), measurements are limited by the small-scale motions in the medium -- turbulence, typically -- that broadens all lines and decreases resonant lines optical thickness.
Quantitative Monte-Carlo simulations help in separating turbulence from resonant scattering \citep[e.g.][]{2013ShangOh, 2013Zhuravleva}. In the perspective of using such a method to measure $H_0$, an important recent observational result is the measurement of resonant scattering in the \ion{Fe}{XXV} He$\alpha$ line in the Hitomi 5\,eV-resolution spectra emitted by the Perseus cluster core \citep{2018HitomiRS}.

%%%%%%%%%%%%
\section{Cluster internal mass distributions}

The hierarchical growth of structure leaves an imprint on the galaxy cluster internal mass distribution. Following this argument, cosmological information must be encoded in the mass profiles and X-ray cluster analyses play an important role in this respect \citep[e.g.][]{2007Buote, 2007Comerford, 2007SchmidtAllen}.
The popular Navarro-Frenk-White (NFW) model \citep{1997NavarroFrenkWhite} for the total matter density distribution $\rho(r)$ in halos has free parameters $r_s$ and $\delta_0$:
\begin{equation}
    \frac{\rho(r)}{\rho_c}  = \frac{\delta_0}{\left( \frac{r}{r_s} \right) \left(1+\frac{r}{r_s} \right)^2}
\end{equation}
This expression may be recast in terms of the concentration $c_{\Delta}=r_{\Delta}/r_s$ by introducing the total mass $M_{\Delta}$ (see Eq.~\ref{eq:mass_delta}) within a given overdensity radius $r_{\Delta}$, providing:
\begin{equation}
    \delta_0 = \frac{\Delta}{3} \frac{c_{\Delta}^3}{\ln(1+c_{\Delta}) - c_{\Delta}/(1+c_{\Delta})}
\end{equation}

The relation between mass and concentration predicted by numerical simulations (and its evolution) is sensitive to the cosmological model.

One expression of this relationship is the dependence of the concentration upon the background density of the Universe at the time the halo collapses.
Using various subsamples drawn from a pool of 44 clusters with deep X-ray observations and assuming the NFW matter distribution for the corresponding halos, \citet{2010Ettori} obtained a set of cosmological constraints.  Such an analysis provides constraints on the combination $S_8 = \sigma_8 \omegam^{\varGamma}$: from 26 clusters with robust measurements, it returns $\varGamma \sim 0.60$ and $S_8 = 0.45 \pm 0.01$, and for a subset of 11 low-entropy core clusters it gives $\varGamma=0.56$ and $S_8 = 0.39 \pm 0.02$. X-ray data plays a key role in this work, as it provides the mass and concentration from the assumption of hydrostatic equilibrium (Eq.~\ref{eq:massTX}).
The likelihood analysis considers incorporating gas fraction measurements in order to break the strong degeneracy along the $\omegam$ dimension. The reported constraints amount to $\sigma_8 = 1.0 \pm 0.2$ and $\omegam = 0.26 \pm 0.02$, the former constraint shifting to $\sigma_8 = 0.83 \pm 0.1$ when only relaxed objects are included. This analysis, albeit providing tight constraints independent from cluster abundance studies, highlights the sensitivity of the method to the model concentration--mass relation and more generally to the model of dark matter distributions as a function of cosmology.

Nevertheless, the NFW parametric profile fails at providing a good fitting functional for halos not entirely relaxed and (or) dynamically disturbed. For those systems, little information is gained from the inferred concentration-mass relation and thus on the underlying cosmological model. Analyses of N-body numerical simulations shows a correlation between the fraction of profiles departing from the NFW model and the cosmological model; this led to introduce `sparsity' as a new, parameter-free, cosmological probe of cluster mass distributions \citep{2014Balmes}. Sparsity $s_{\Delta_1, \Delta_2}$ is defined as the ratio of masses computed at two different overdensities chosen to encompass the intermediate- to large-radial range: $s_{\Delta_1, \Delta_2} = M_{\Delta_1}/M_{\Delta_2}$ with $\Delta_1 < \Delta_2$. Typically, $\Delta_1 \sim 200$ and $\Delta_2 \sim 500$, but other radii are sensible choices as long as $\Delta_1 \gtrsim 100$ and $\Delta_2 \lesssim 2000$.
Simulations indicate small scatter in this value and importantly, small percent-level mass dependence \citep{2019CorasanitiRasera}; even at $z \sim 1$ \citep{2018Bartalucci}.
Using this property and knowing the predicted mass function at two different overdensity radii and at a given redshift bridges sparsity and the model cosmology \citep{2014Balmes, 2018Corasaniti}:
\begin{equation}
    \int_{M_{\Delta_2}^{\rm min}}^{M_{\Delta_2}^{\rm max}} \frac{\dd n}{\dd M_{\Delta_2}} \dd \ln M_{\Delta_2} = \langle s_{\Delta_1, \Delta_2} \rangle \int_{\langle s_{\Delta_1, \Delta_2} \rangle M_{\Delta_2}^{\rm min}}^{\langle s_{\Delta_1, \Delta_2} \rangle M_{\Delta_2}^{\rm max}} \frac{\dd n}{\dd M_{\Delta_1}} \dd \ln M_{\Delta_1}
\end{equation}
where the integration runs over the mass range of the cluster sample and brackets indicate the average across the sample. Various estimators of this average exist for a given sample; any inconsistency among estimates reveals possible systematic biases in the determination or in the extrapolation of mass profiles \citep{2019CorasanitiRasera}.
Application of the sparsity test to a sample of 104 clusters with masses estimated from X-ray data assuming hydrostatic equilibrium leads to $\omegam=0.42 \pm 0.17$ and $\sigma_8=0.80 \pm 0.31$ \citep{2018Corasaniti}; degeneracy between parameter constraints makes $S_8 = \sigma_8 \sqrt{\omegam} = 0.48 \pm 0.11$ more relevant an output of this test. This result takes advantage of an increased sensitivity of the sparsity test at around $z \sim 0.5$, as shown by analyses of a large mock catalogue of halos demonstrating the ability of the method to recover input cosmological parameters. Various systematic biases under scrutiny are found to have mild impact on the results; most prominent are selection effects and a radial-dependent hydrostatic mass bias. Hydrodynamical simulations incorporating the effect of baryons indeed indicate a few percent impact from the latter. Finally, the results are sensitive to the choice of a theoretical mass function and the previously quoted constraints are slightly offset by a change in this assumption.

\section{Pink elephants}

According to the scenario for structure formation depicted in Sect.~\ref{sect:basic_cosmology}, we do not expect massive clusters to have formed at high redshifts. Because cluster detection signals scale positively with mass, especially at X-ray wavelengths, an extreme-mass system should easily be discovered in survey data if it exists. Exceptional massive structures therefore offer a promising test of the concordance model\footnote{Or variations thereof, in particular non-Gaussianities in the primordial fluctuation field, see also note~\ref{note:planckNG}. By condordance model, we mean the most commonly accepted cosmological model, such as that depicted in Sect.~\ref{sect:basic_cosmology}.}. Similarly as for cluster count studies, a number of effects and systematics must be understood before concluding to falsification of the cosmological model; among them are the uncertainties in the theoretical mass function \citep[e.g.][]{2012HolzPerlmutter}, the mass calibration and the scatter in the mass-observable relation \citep[e.g.][]{2011Mortonson, 2012HarrisonColes}, especially its impact on cluster selection \citep[e.g.][]{2012Hoyle}.

The recent discovery of massive high-redshift systems in wide surveys triggered an increase of interest in this area of research. Among others we may cite XMMU~J2235 at $z=1.393$ \citep{2005Mullis}, RDCS1252.9 at $z=1.24$ \citep{2005Lombardi} or SPT-CL~J2106-5844 at $z=1.132$, the most massive $z>1$ cluster found in the early SPT survey \citep{2011Foley}, with $M_{200c} \sim 10^{15} M_{\odot}$. Abundant literature followed these discoveries, feeding a debate on their `rareness' in the context of the standard model and its extensions. As of today, it seems none of these systems indicates clear tension with $\varLambda$CDM. For the purpose of this Chapter, we follow \citet{2013HarrisonHotchkiss} in classifying statistical tests into two categories: tests relying on extreme value statistics and those based on the rareness of a given cluster of a certain mass and redshift in $\varLambda$CDM. A discussion of extreme pairwise velocities is appended, that comes as an interesting complement to these tests.

This section entails studies of galaxy clusters in X-rays, although the search for extreme objects represents a wide field of investigations in cosmology, involving structures of many kinds: walls, voids, superclusters \citep[e.g.][]{2011ShethDiaferio, 2012Park, 2016Sahlen}.

%%%%%%%%%%%% Class I tests: Extreme value statistics
\subsection{Extreme-value statistics}

We may examine the statistics of extrema in a random field smoothed with a kernel of a given scale within a volume of the Universe (a `patch') of a given size \citep{2011Colombi}.
The probability distribution of the highest peak equals the probability $P_0(\nu)$ that no local maximum with normalized density $\nu_{\rm max}$ exceeds a certain threshold value $\nu$ \citep[e.g.][]{1992Bertschinger}:
\begin{equation}
    {\rm Prob.}(\nu_{\rm max} < \nu) = \int_{-\infty}^{\nu} p_G(\nu_{\rm max}) \dd \nu_{\rm max} = P_0(\nu)
\end{equation}
where $p_G(\nu_{\rm max})$ denotes the distribution of local maxima among patches; this is the statistics of interest here, easily obtained with $p_G = \dd P_0/ \dd \nu$.
The `count-in-cell' formalism allows further progress. One has \citep[e.g.][]{1979White}:
\begin{equation}
\label{eq:p0_colombi}
    P_0(\nu) = \exp \left[ -n(> \nu) V \sigma(N_c) \right]
\end{equation}
with $N_c \equiv n(>\nu) V \bar \xi^{h}_2$.
The expression for $N_c$ comprises the average number of peaks above threshold, $n(>\nu)$, computed over the volume of interest $V$; clearly $N_c$ represents the typical number of halos exceeding $\nu$ in an overdense patch.
In general $\sigma$ is a function involving all the halo correlation functions $\bar \xi^h_N$ of order $N$, averaged over the volume $V$. A simple formula is obtained in the limit of massive peaks ($\nu \gg 1$) and large separations \citep[e.g.][]{1999Bernardeau}:
\begin{equation}
    \label{eq:sigma_colombi}
    \sigma(y) = \left( 1+ \frac{\theta}{2} \right) {\rm e}^{-\theta}, \text{\ with \ } \theta {\rm e}^{\theta} = y.
\end{equation}
Releasing some of these assumptions\footnote{It is beyond the scope of this Chapter to review the assumptions leading to this formula, the interested reader may refer to \citet{2011Colombi}.} allows to enlarge the range of applicability of this statistics to smaller patches, smaller masses and smaller separations between halos.
An interesting limiting case is:
\begin{equation}
\label{eq:sigma_smallNc}
    \sigma(N_c) \simeq 1- N_c/2   \ \ \ \ (\mathrm{if}\ N_c \ll 1)
\end{equation}
For $N_c=0$, the set of equations presented above corresponds to the so-called `Poissonian' form. An evaluation of its validity is carried in \citet{2011Davis}, both analytically and with the help of a numerical experiment using an N-body simulation. The analytic comparison with the complete formula involving halo clustering (Eqs.~\ref{eq:p0_colombi} and~\ref{eq:sigma_colombi} and invoking a linear halo bias $b$) shows good agreement for patches of size $L\sim 100 h^{-1}$~Mpc. The comparison with N-body simulated patches demonstrates good performance of the full formula for patches of sizes above $\sim 20 h^{-1}$~Mpc. In all cases the agreement is better at the high-mass end of the distributions (large $\nu$ and large separations).

The expression in the `Poissonian' limit is exploited by \citet{2011Waizmann} with the specific goal of relating the shape of $p_G(m)$ to the following parametric form:
\begin{equation}
    \label{eq:gev_statistics}
    p_{\rm GEV}(u) = \exp \left[- (1+\gamma z)^{-1/\gamma} \right], {\ \ \ } z=(u-a)/b
\end{equation}
with free parameters $a, b, \gamma$. This distribution is sometimes called `General Extreme Value statistics' (GEV). With an assumption for the expression of the halo mass function $n(>m)$ \citep[e.g.][]{1999ShethTormen, 2008Tinker} and by performing first-order Taylor expansion of both $p_G$ and $p_{\rm GEV}$ around their maximal values \citep[see e.g.][for a derivation]{2011Davis}, one obtains analytic expressions for the parameters $(a, b, \gamma)$, with a dependence on cosmology and on the growth of structure. In particular, the study shows that the most-likely extreme halo mass in a patch of sky, directly related to $a$, is a sensitive probe of the $\varLambda$CDM model. To demonstrate this result, \citet{2011Waizmann} recover $a$ from a fit of the full cumulative distribution of extreme masses extracted from $\sim 500$ independent Universe patches at $1 < z < 1.5$ with a limiting sensitivity $M \sim 10^{14.5} h^{-1} M_{\odot}$. The proposed cosmological test therefore consists in measuring the mass of the most extreme cluster in each patch with a certain precision, then group clusters in mass bins to form a cumulative distribution and finally fit the GEV formula (Eq.~\ref{eq:gev_statistics}) to retrieve the best-fit $a$ and use its value to discriminate between various dark energy models or gravitation theories.

In an investigation on constraints on non-Gaussianities, \citet{2012ChongchitnanSilk} provide an independent and detailed derivation of the extreme value statistics for halos above a given mass threshold (Eq.~\ref{eq:p0_colombi}, \ref{eq:sigma_colombi}); of particular interest is their analytic development carried for weak non-Gaussianities of the underlying field, allowing to expand $\sigma$ one order beyond the approximation of Eq.~\ref{eq:sigma_smallNc}. The probability that XMMUJ0044.2 \citep{2011Santos} is the most massive system in the XDCP survey is consistent with $f_{\rm NL} =0$; the usual degeneracy between $\sigma_8$ and $f_{\rm NL}$ requires fixing its value to that of WMAP-7. The assumptions made on the shape of the mass function play an important role in this analysis, all the more at higher redshift (XMMUJ0044.2 has redshift $z=1.579$). This work does not involve fitting or asymptotic formula such as Eq.~\ref{eq:gev_statistics}. However the convergence of the distribution towards this family of functions is found to be insensitive to the value of $f_{\rm NL}$.

%%%%%%%%%%%%%%% Class II tests: rareness
\subsection{Rareness of events}

One may ask the question of whether a particular system is the most massive in a given survey. For instance, \citet{2011Cayon} examine the case of XMMU2235 at $z\sim 1.4$. Poisson statistics serve in deriving the posterior probability for $f_{\rm NL}$ and $\sigma_8$, using a mass function modified to account for local non-Gaussianities. The study examines various systematics including Eddington bias, assumptions on the concentration parameter involved in the mass determination and triaxiality as well as assumptions on the theoretical mass function. This work finds a tentative positive value for $f_{\rm NL}$ (see Sect.~\ref{sect:clusters_clustering} for a definition of this parameter), although zero is not excluded.

As pointed out by several authors \citep[e.g.][]{2011Hotchkiss, 2013HarrisonHotchkiss} some of the statistics quantifying rareness of a single cluster found at a given mass and redshift suffer from a bias due to their neglecting the full population of extreme clusters of different masses and redshifts. Low probability values delivered by a set of `exclusion curves' \citep[as proposed in e.g.][]{2011Mortonson} do not necessarily point to \emph{exceptional} systems after accounting is made for the underlying distribution of halos in the mass--redshift space. However such tabulated probabilities proved useful in estimating the probability that no cluster is more massive and more distant than a given object \citep[e.g.][]{2011Jee}. In this regard, rareness of a sample of clusters does not imply falsification of the $\varLambda$CDM model: an instructive critical view may be acquired with the help of Monte-Carlo numerical simulations \citep{2012Hoyle}. For instance, \citet{2019Kim} provide a new estimate of the mass of SPTJ2106; using one of the rareness statistics of \citet{2011Hotchkiss} their study assesses a discovery probability of $\sim 72\%$ within the survey volume. This value is higher than the initial estimate \citep[$\sim 7\%$,][]{2011Foley} thanks to their updates of the mass measurement and of the statistical treatment.
An additional rareness study of XMMUJ0044.2 is described by \citet{2015Tozzi} following deep \emph{Chandra} X-ray observations and a new estimation of its (hydrostatic, isothermal) mass; the authors consider its existence as an additional hint tentatively indicating tension in the standard $\varLambda$CDM model or a significantly higher value for $\sigma_8$ than found by WMAP-7.

\subsection{Extreme pairwise velocities}

A complementary test involves pairwise velocities in major cluster mergers, whose extreme amplitudes are predicted as exceptional in the concordance model.
Current X-ray observatories can provide information on the shock velocity for collisions near the plane of sky \citep[e.g.][]{2002Markevitch, 2016Dasadia}.
Observational searches at CCD spectral resolution with \emph{Chandra} hints towards a tentative detection of bulk motions perpendicular to the plane of sky in the gas of the Bullet Cluster \citep{2015LiuYuTozzi}, however robust velocity measurements require high-resolution X-ray spectroscopy \citep[as in][]{2018HitomiPerseus}, a technique soon to be routinely available with modern X-ray observatories \citep{2013NandraAthena, 2020XRISMScience}.
The Bullet Cluster (1E0657-56) is one cluster concentrating most of attention given its well measured properties, in particular the shock velocity ($v_s \simeq 4700$ km s$^{-1}$). However the shock speed might be higher than the subcluster speed, the latter amounting to $\sim 3000$\,km\,s$^{-1}$ \citep[e.g.][]{2007Springel}, hence lowering the exceptional character of this system. Aided with a large N-body simulation \citet{2014Watson} found no inconsistency with the $\varLambda$CDM model, although this study seems to be missing a complete account for systematics. The simulation-based work in \citet{2015Bouillot} also concludes to an absence of inconsistency, complementing findings from independent studies \citep[e.g.][]{2015KraljicSarkar, 2015LageFarrar}.
The debate is still open though, as many more extreme systems such as the Bullet cluster would seriously challenge the standard model. The `El Gordo' system, a galaxy cluster merger at $z=0.87$, might be one such object that is raising an inconsistency given its mass $M_{200} = (2-3) \times 10^{15}\,M_{\odot}$ and infall velocity $V_{inf} \sim 2500$\,km\,s$^{-1}$ \citep{2021Asencio}.

%%%%%%%%%%%%%%%%%%%%%%%%%%%%%

\section{Clusters as gravitational theory probes}

The scenario for structure formation in the $\varLambda$CDM model described in Sect.~\ref{sect:basic_cosmology} puts gravity at the centre of the play. The distribution, appearance, properties of galaxy clusters therefore strongly depend on the underlying law of gravity; detailed studies offer tests of the gravitation theory at scales between cosmological and galactic. Modifications to the General Relativity have an impact on the abundance of clusters, on their shape, on the gravitational potential they live in and on virtually any observational feature we have discussed in this Chapter.
We refer the reader to recent reviews \citep[e.g.][]{2012KravtsovBorgani, 2015deMartino, 2016Koyama, 2018CataneoRapetti, 2019Ferreira} for a comprehensive overview and results from tests of gravity including galaxy clusters. In this Chapter we briefly describe three example studies specifically relying on X-ray cluster observations and samples.

As mentioned in previous sections, experiments probing the abundance of X-ray galaxy clusters may constrain the exponent $\gamma$ parametrizing the linear growth-rate $f(z) \simeq g(z)\equiv \omegam(z)^{\gamma}$, thereby offering a (so far successful) consistency check of General Relativity \citep[e.g.][]{2009Rapetti, 2010Rapetti, 2013Rapetti}.
Constraining alternative theories of gravitation necessitates a full understanding of the non-linear growth of structures and halo collapse in these frameworks. Analytic calculations provide insight into the scales and mass regimes most affected by such modifications \citep[e.g.][]{2012BraxValageas}. N-body simulations complement these works and provide fits to the modified halo mass function, for instance in Hu-Sawicki $f(R)$ gravity and Dvali-Gabadadze-Porrati models \citep[e.g.][]{2008Oyaizu, 2011Ferraro, 2021Gupta}. The reanalysis of the abundance of X-ray cluster samples puts constraints on the absolute value of the perturbation parameter $f_{R0}$ entering one specific model of $f(R)$ modified gravity, to less than $10^{-4.79}$ (at 95\% confidence level) \citep{2009Schmidt, 2015Cataneo}. A value of zero for $f_{R0}$ indicates consistency with $\varLambda$CDM, while deviations from this model are of the order of this perturbation parameter. Systematics for this class of tests of modified gravity are related to uncertainties on the halo mass function, cluster selection effects and the mass-observable link, similarly as discussed in Sect.~\ref{sect:cosmo_abundances}.

A distinct use of clusters for testing gravity relies on their matter distribution. Following the proposal of \citet{2017Verlinde} for a theory of `Emergent Gravity', a number of analyses concentrated on comparisons of the observational X-ray properties of galaxy clusters to the radial distribution of the apparent dark matter. The latter arises in the theory in a form that depends only on the baryonic mass (predominantly as X-ray emitting gas) and $H_0$. Whether estimating total masses from hydrostatic equilibrium equations \citep{2017Ettori} or weak-lensing measurements \citep{2019Tamosiunas, 2019ZuHone} of several tens of systems, no preference is found for such an alternative gravitation theory, at least in its current form.

The gravitational potential may be probed through the gravitational redshift experienced by photons losing energy as they escape deep potential wells. The shape of the potential is changed as gravity is modified. Stacking together the spectroscopically measured galaxy velocities in a large sample of optical clusters, \citet{2011Wojtak} identified a shift of velocity for satellite galaxies relative to centrals, indicative of the sought-after gravitational redshift signal. Further analyses incorporating all relevant relativistic effects and consistent changes in the halo mass function however conclude to little discriminating power of this technique if only dynamical mass information for a narrow range of cluster masses is available \citep[e.g.][]{2013Zhao}. A study using X-ray-based masses obtained for $\sim 2500$ galaxy clusters detected in the SPIDERS survey also concludes to no preference between General Relativity and a special case of $f(R)$ gravity \citep{2021Mpetha}. In all these studies, the amount of the measured velocity shift is of order $10-20$\,km\,s$^{-1}$. The future use of high-resolution spectroscopic X-ray data may probe gravitational redshifts of the gas instead of individual galaxies \citep[e.g.][]{2000Broadhurst, 2017Sakuma}. This observational technique provides a reduction of the noise inherent to measurements due to the high number of gas parcels in comparison to galaxies, at the expense of finely controlling all effects creating $\sim$ few 10\,km\,s$^{-1}$  line shifts in the spectral emission (such as turbulent and bulk motions). Only with upcoming instrumentation will this observational technique be accessible, together with the observational and theoretical improvements in our understanding of the intra-cluster gas dynamics and chemical enrichment.

\section{Selection function}

When the catalogs of galaxy clusters are derived using X-ray data, X-ray properties of clusters and methods of cluster detection form the selection function of the survey. The shape of the X-ray emission of the cluster in combination with the detection method affects the selection: unresolved X-ray emission (typically covering $2-6^\prime$ radius) formed the basis for the bulk of cluster catalogs based on \emph{ROSAT} All-Sky survey data, which for distant clusters covers most of the cluster, while for the nearby ($z<0.1$) clusters it covers only the core, where X-ray emission scales poorly with cluster mass. While accounting for the total flux of the cluster is performed, such selection leads to unbalanced representation of merging and relaxed clusters. Another drawback of this approach is that 90\% of such sources are not clusters and one has to clean for contamination in the sample and in cluster X-ray properties. A huge improvement in the purity of cluster selection came with using the extent of X-ray emission. The selection function in those surveys was found to depend on the cluster core radii. In order to assess the limitations of the complexity of the cluster shapes, cloning of observed clusters has been used to test the cluster detection algorithms. 

Scatter in the cluster shape has been found to correlate with scatter in cluster X-ray luminosity, which affects cluster selection \citep{2019Kaefer}. The full account for covariant scatter on cluster selection is given in \citet{2020Finoguenov}. 
In terms of the selection, the best area of the cluster to use seems to be between 1/4 and 1/2 of $R_{500c}$, and efforts have been made to create the detection procedures for that \citep{2020Eckert}. Avoiding cluster cores in the detection has been widely used in deep X-ray surveys \citep{2010Finoguenov, 2015Finoguenov}.
The advantage one uses in those methods is the large information on clusters shapes available at X-rays.

In the forward modelling, for each prediction of mass function, one predicts the full set of cluster X-ray mass proxies and mimics the selection process. Detection of galaxy clusters forms the selection function, while the procedure of measurement of the cluster properties forms the cluster distribution function. On the procedure of measurements of cluster properties, there is a large variety of approaches taken. Even to obtain the cluster $L_X$ one has to determine the cluster extent. However, in most cases only the central part of the cluster is detected and extrapolation of cluster properties is required. For that, one either uses a beta profile, which is either measured \citep{1999Vikhlinin} or is taken from scaling cluster libraries \citep{2020Finoguenov}, or sets the definition of cluster $L_X$ to be within the zone of likely detection (e.g. 300\,kpc for XXL). REFLEX and NORAS surveys \citep{2004Boehringer} do not have a scalable definition of $L_X$ and corrections have been computed for those clusters \citep{2010Mantz, 2011Piffaretti}.

The purpose of deriving a selection function is to emulate the behaviour of a detection chain that led to the construction of a sample, while varying assumptions on the physics and cosmological model. Expensive end-to-end simulations are an option which accounts for many of the details \citep[projection effects, point-source contamination, etc., see e.g.][]{2020Comparat}, while scaling models are an alternative that bypasses the expensive computations, but need to formulate strong assumptions on their behaviour. CODEX, XXL selection functions are somehow half-way between these two extremes. In order to make sure that certain cluster shapes are not missed in the current X-ray libraries, studies of clusters selected using different ways (e.g. SZ effect, galaxy concentrations or lensing) are used. The most obvious result of those searches so far is finding that clusters without the cool core are more abundant, compared to the results of REFLEX/NORAS catalogs. Similar results are obtained using spatially resolved X-ray searches for clusters.

In most cases, identification of galaxy clusters requires detection of cluster {\it galaxies}. Limitations of these data will affect the selection of clusters and of importance is the covariance of optical and X-ray properties of the sample. The current results \citep{2019Farahi} also suggest the anti-covariance, which implies that cluster entering the sample by upscatter in X-ray properties, might leave the sample due to downscatter in optical richness. The scatter of properties in clusters is linked to the physics of clusters. X-ray luminous clusters have bright cores and are preferentially relaxed. Merger clusters tend to have high richness \citep{2019Mulroy}. The study of \citet{2019Klein} has introduced a concept of using cluster identification to study the chance association between the X-ray source and the optical counterpart. Using an overlap with Klein et al. sample, \citet{2020Finoguenov} concluded that chance identification starts to be important below the 10\% completeness boundary of the X-ray survey.

\section{Conclusions and forward look}

X-ray clusters characterize well the massive end of halo mass function, and there are no limitation imposed on future studies to extend the X-ray cluster searches to redshifts beyond the current record of 2.5. The future studies will benefit from improved sensitivity in the soft X-ray band (0.1-2 keV) (e.g. of \emph{Athena}) and spatial resolution better than a few arcseconds, as anticipated from the {\it Lynx} mission. Large catalogs of clusters delivered by eROSITA all sky will serve a great input for cosmological studies. The best behaved part of the X-ray emission of clusters is located between 0.2 and $R_{500c}$, where X-ray (which is necessarily a surface brightness) selection have best chances to be complete. 

Currently, $z<0.5$ X-ray cluster studies have reached the limit of systematics. This requires both mass calibration and improved modelling of all baryonic cluster components. 
A variety of shapes of X-ray emission of clusters might be associated with a difference in the cluster baryonic content induced by AGN feedback. In order to understand that, detailed X-ray studies are required to recover the diversity of gas mass fraction close to the virial radius. 
Historically, these searches have been hampered by problems with robust definitions of massive halos using optical cluster catalogs and deficiencies in existing X-ray catalogs at low-z. A number of Large Programs on \emph{XMM-Newton} are currently underway to progress on this issue. 
Deep X-ray and spectroscopic surveys have also been proven to be the best way to address this \citep{2009Finoguenov, 2012Connelly}. A special topic is whether the distribution of X-ray properties of clusters follows the power law dependence on mass with a log-normal scatter, or has a more complicated functional form. With increasing size of X-ray cluster samples, and increasing variety of cluster detection methods, one expects to collect more data on this subject.

Variety of baryonic properties of halos also affects weak lensing cosmology and present a new research focus on X-ray groups and clusters. 
For instance the high precision of \emph{Euclid} optical and near-infrared measurements will be sensitive to detailed processes associated with galaxy formation, including the growth of black holes and their feedback (co-evolution). \citet{2011Semboloni} predict the importance of the contribution of galaxy groups to the shear power spectrum down to  $10^{13} M_\odot$.  \citet{2020Debackere} confirm this finding through modelling the distribution of baryons in groups following observations of stellar and hot intragroup medium. Baryonic feedback processes, either from supernovae (SNe) or active galactic nuclei (AGN) redistribute the gas, which changes the matter power spectrum, in particular on scales where the sensitivity of cosmic shear is maximal. \citet{2011Semboloni} showed that ignoring this effect leads to large biases, whereas marginalising over current model uncertainties weaken the constraining power of weak lensing shear experiments. The way forward is to constrain feedback directly and update its implementation in models \citep[e.g.][]{2017McCarthy}. Groups are the best targets for resolving this issue because on their scale: on one hand the dominant feedback mechanism must change from stellar processes, such as galactic winds to AGN-driven processes and imprint telltale signatures on the warm gas, as the binding energy and the AGN output are similar; on the other hand, all the components of galaxy groups can be robustly determined and groups of galaxies are the lowest mass structures that impact the shear signal on the scales of interest for \emph{Euclid} cosmology.

\section{Resources}

Any initial analysis of X-ray observations is performed with a mission-dependent software package, with a common goal of producing calibrated event list and generating auxiliary responses. For a collection of the software on various mission we would like to point the reader to HEASARC\footnote{\url{https://heasarc.gsfc.nasa.gov/}}. Most of the data products has been stored using FITS format with the FTOOLS software, available on the same site, as well as a package for spectral analysis, XSPEC. Detailed study of line emission shall benefit the largest database of lines to date, available through another spectral fitting code, SPEX\footnote{\url{https://www.sron.nl/astrophysics-spex}}. 

Cosmological codes relevant to galaxy clusters are available through Astropy\footnote{\url{https://docs.astropy.org/en/stable/cosmology/index.html}} \citep{2013Astropy}
and COLOSSUS\footnote{\url{http://www.benediktdiemer.com/code/colossus/}} \citep{2018Diemer}. Cosmology codes in FORTRAN and other languages are available from E.~Komatsu's website\footnote{\url{https://wwwmpa.mpa-garching.mpg.de/~komatsu/codes.html}}.
A cosmic variance python module \citep{2019LacasaGrain} is available online\footnote{\url{https://github.com/fabienlacasa/PySSC}}.

\bibliographystyle{aa}
\bibliography{bibliography}

\end{document}